\documentclass[12pt]{article}
\usepackage{graphics,cite}
\usepackage{epsfig}

\def\beq{\begin{equation}}
\def\eeq{\end{equation}}
\def\beqa{\begin{eqnarray}}
\def\eeqa{\end{eqnarray}}
\def\bet{\begin{tabular}}
\def\eet{\end{tabular}}

\newcommand{\sect}[1]{\setcounter{equation}{0}\section{#1}}

\newcommand{\msbar}{$\overline{\mbox{MS}}$ }

\def \lsim
{\raisebox{-3pt}{$\>\stackrel{<}{\scriptstyle\sim}\>$}}

\begin{document}

\begin{titlepage}

\begin{flushright}
ROME1/1436/06
\end{flushright}

\vspace{1.3cm}

\begin{center}
{\Large \bf Modelling  non-perturbative corrections\\
to bottom-quark fragmentation\\}
\end{center}

\vspace{5mm}

\begin{center}
{\large \bf Ugo Aglietti$^{1,2}$, Gennaro Corcella$^1$\\ 

\vspace{0.2cm}
and Giancarlo Ferrera$^{1,2}$}\\

\vspace{5mm}

{$^1${\sl Dipartimento di Fisica, Universit\`a di Roma `La Sapienza'}\\ 
{\sl P.le A.~Moro~2, I-00185 Roma, Italy}}

\vspace{3mm}

{$^2${\sl INFN, Sezione di Roma}\\
{\sl P.le A.~Moro~2, I-00185 Roma, Italy}}

\end{center}

\vspace{1cm}

\begin{abstract}
We describe $B$-hadron production in $e^+e^-$ annihilation at the
$Z^0$ pole by means of a model including non-perturbative corrections
to $b$-quark fragmentation as originating, via multiple
soft emissions, from an effective QCD coupling constant,
which does not exhibit the Landau pole any longer and includes
absorptive effects due to parton branching.
We work in the framework of perturbative fragmentation functions 
at NLO, with NLL DGLAP evolution and
NNLL large-$x$ resummation in both coefficient function 
and initial condition of the perturbative fragmentation function.
We include hadronization corrections via the
effective coupling constant in the NNLO approximation and do not add
any further non-perturbative fragmentation function.
As part of our model, we perform the
Mellin transforms of our resummed expressions exactly. 
We present results on the energy distribution of $b$-flavoured hadrons, 
which we compare with LEP and SLD data, in both $x$- and $N$-spaces.
We find that, within the theoretical uncertainties on our calculation,
our model is able to reasonably reproduce the data at $x\lsim 0.92$ and
the first five moments of the $B$ cross section.

\end{abstract}

\end{titlepage}

\sect{Introduction}
\label{intro}

We study $b$-quark fragmentation and $B$-hadron production
in $e^+e^-$ annihilation using a model which implements
power corrections via an analytic effective coupling constant.
Such a model will allow us to make hadron-level predictions, directly
comparable with the accurate experimental data from the SLD \cite{sld} and 
LEP \cite{aleph,opal,delphi} collaborations.

Let us describe in physical terms the process we are interested in:
\beq
\label{proc_in}
e^+ e^- \, \to \, Z^0 \, \to \, B \, + \, X_{\bar b}, 
\eeq
where $X_{\bar b}$ is a generic hadronic final state with a $\bar b$ quark.
The corresponding parton-level process, assuming that
an arbitrary number of gluons is radiated, reads:
\beq\label{bbg1}
Z^0 \, \to \, b \, \bar{b}\to b\, \bar b\, g_1\,\dots\, g_n.
\eeq
In the $Z^0$ rest frame, the 
$b$ and the $\bar b$ are initially emitted back-to-back
with a large virtuality, of the order of the hard scale $Q= m_Z$.
Then, they reduce their energy by emitting gluons,
mainly of low energy or at small angle (soft or collinear radiation).
As long as the virtuality of the $b$ quarks is large enough, 
we can neglect the $b$ and $\bar{b}$ masses. 
Considering, for simplicity, the case of one single emission,
\begin{equation}
e^+ e^- \, \to \, Z^0(q) \, \to \, b(p_b) \, 
+ \, \bar{b} (p_{\bar{b}}) \,  + \, g(p_g),
\end{equation} 
and defining the energy fractions
\begin{equation}
x_i \, \equiv \, {{2p_i\cdot q}\over {m_Z^2}} \, = \, \frac{2E_i}{m_Z},
\end{equation}
where $i \, = \, b,\bar{b},g$ and $x_b+x_{\bar b}+x_g=2$, a straightforward
kinematical computation yields:
\begin{equation}
1-x_b={{x_{\bar b}x_g}\over 2}(1-\cos\theta_{\bar b g}),
\label{xbb}
\end{equation}
with $\theta_{\bar bg}$ being the angle between the gluon and the $\bar b$.
From Eq.~(\ref{xbb}), we see that collinear
emission from the $\bar b$ ($\theta_{\bar bg}=0$) implies $x_b$=1. 
Likewise, by crossing $b\to\bar b$, one can show that 
collinear radiation off the $b$, i.e. $\theta_{bg}=0$, corresponds to
$x_{\bar b}=1$.
Soft-gluon radiation ($x_g=0$) yields 
$x_b=x_{\bar b}=1$.
We can identify two main different
mechanisms of energy loss of the $b$: a direct loss,
related to soft or 
collinear emissions off the $b$ itself, and an indirect loss,
when the gluons are radiated off the $\bar b$ (hard and large-angle emissions
are not enhanced).
The asymmetry between the $b$ and the $\bar{b}$ jets
is not dynamical, but is just related to the fact that, e.g., we decided to
measure the energy of the $b$, and not that of the $\bar{b}$.

When the virtuality of the $b$ has reduced to a value of the order of its mass,
mass effects become substantial, and 
when it becomes comparable with the hadronic scale, 
${\cal O}(1~\mathrm{GeV})$, the $b$
approaches the non-perturbative phase of its evolution. 
The hadronization, for example, into a $B$ meson can be described as
due to the radiation of a light 
$q\bar{q}$ pair, and the subsequent combination of the $b$ and the $\bar q$ 
to form a $B \,= \, (b\bar{q})$ state:
\beq
\label{mesone}
b \, \to \,  B\, + \, q.
\eeq
In the process (\ref{mesone}), the colour
of the heavy $b$ is coherently transferred to the light $q$; 
we shall assume that no more radiation 
is emitted after the colourless hadron has been formed.
The hadronization transition (\ref{mesone}) is sensitive to 
non-perturbative power corrections $\Lambda/m_b$, with $\Lambda$ being the QCD 
scale, e.g., in the \msbar scheme. Such power corrections - for the time
being - cannot be calculated 
from first principles, but are usually described 
by means of phenomenological hadronization models, such as the 
Kartvelishvili \cite{kart}
or the Peterson \cite{pet}
models, containing parameters which are to be fitted to experimental data.
In Ref.~\cite{gardi}, furthermore, non-perturbative
effects in bottom fragmentation at large $x$ were described as a 
shape function of $m_b(1-x)$ and implemented using Dressed 
Gluon Exponentiation \cite{gardi2}.

In this paper, we shall follow a different approach to model power
corrections to $b$-quark fragmentation. 
We shall assume that the transition (\ref{mesone})
can be described in terms of an effective analytic coupling
constant $\tilde\alpha_S$, which
incorporates non-perturbative power corrections and,
unlike the models above mentioned, does not present any free parameter to be
fitted to data. 
Within our model, whenever we use $\tilde\alpha_S$ instead of the
standard $\alpha_S$, 
the energy of the $b$ quark will be identified
with the one of the observed $B$ meson, i.e. $E_b\to E_B$.
The non-perturbative model which we shall use hereafter
was already used in \cite{noi4} in the context of 
$B$-meson decays and interesting results were found.
The comparison with the data on the photon spectrum
in radiative decays and on the hadron-mass distribution in
semileptonic decays showed good agreement, while discrepancies were
found for the purpose of the electron energy distribution.

An analogous model can be introduced for the fragmentation
of a $b$ into a baryon, such as, for example, a hyperion $\Lambda_b$. 
The $b$ quark emits two light $u\bar{u}$ 
and $d\bar{d}$ pairs and combines with the $u$ and the $d$  
to form the colourless $\Lambda_b \, = \, (bud)$:
\beq
\label{iperione}
b \, \to \, \Lambda_b \, + \, \bar{u} \, + \, \bar{d},
\eeq
where the $\bar{u}\bar{d}$ system is emitted with the same
colour state as the initial $b$.
The process (\ref{iperione}) will be also described by means of 
an effective coupling constant: however, 
there is no reason {\it a priori} to assume 
that the effective coupling constant
for (\ref{iperione}) is equal to that for (\ref{mesone}).

Turning back to the parton-level process (\ref{bbg1}),
the differential cross section has a
perturbative expansion containing large logarithms of the form:
\beq
\label{masslogs}
~~~~~~~~~~~~~~~~~~~~~~~~~~~~~~~~~~~~~~~~~~~~~~~~~~~~~
\alpha_S^n \, \ln^k \frac{m_Z^2}{m_b^2} 
~~~~~~(n=1,2,3\cdots\infty,~k = 1,2,\cdots,n), 
\eeq
which are related to collinear emissions
of partons with transverse momenta ranging between scales fixed by the
particle masses:
\beq
\alpha_S \, \ln \frac{m_Z^2}{m_b^2}
 \, = \,
\alpha_S \, \int_{m_b^2}^{m_Z^2} \frac{dk_{\perp}^2}{k_{\perp}^2}.
\eeq
At higher orders, the leading contributions, 
$k \, = \, n$ in Eq.~(\ref{masslogs}), 
are associated with phase-space regions ordered in transverse momentum:
\beq
\label{ktordered}
k_{\perp 1} \, > \,  k_{\perp 2} \, > \cdots \, > \,k_{\perp n},
\eeq
leading to integrals of the form
\beq
\label{leadinglo}
\alpha_S^n 
\, \int_{m_b^2}^{m_Z^2} \frac{d k_{\perp 1}^2}{k_{\perp 1}^2}
\, \int_{m_b^2}^{k_{\perp 1}^2} \frac{d k_{\perp 2}^2}{k_{\perp 2}^2}
\, \, \cdots \,
\, \int_{m_b^2}^{k_{\perp n-1}^2} \frac{d k_{\perp n}^2}{k_{\perp n}^2}
\, = \, \frac{1}{n!} \, \alpha_S^n \, \ln^n  \frac{m_Z^2}{m_b^2}.
\eeq
Hence, in the first-order cross section,
the collinear logarithm exponentiates.

The $b$-quark spectrum is also affected by large logarithms
\beq
\label{xlogs}
~~~~~~~~~~~~~~~~~~~~~~~~~~~~~~~~~~~~~~~
\alpha_S^n \, \left[ \frac{\ln^k(1-x)}{1-x} \right]_+
~~~~(n = 1,2\cdots,\infty,~k = 0,1,2,\cdots, 2n-1),
\eeq
also called threshold logarithms, 
which are enhanced for $x\to 1$, i.e. for soft or collinear emission.
The plus distributions in Eq.~(\ref{xlogs}) are defined as:
\beq
\left[ \frac{\ln^k(1-x)}{1-x} \right]_+
\, \equiv \, \lim_{\epsilon \to 0^+} 
\, \left[
\Theta(1-x-\epsilon)\frac{\ln^k(1-x)}{1-x} 
\, - \, \delta(1-x-\epsilon) \, \int_0^{1-\epsilon} \frac{\ln^k(1-x')}{1-x'} \, dx'
\right].
\eeq
The contributions in (\ref{xlogs}) originate 
from the following double-logarithmic integrals:
\beq
\left[ \frac{\ln(1-x)}{1-x} \right]_+
\, = \, 
- \, \left[ \int_0^1 \, \int_0^1 \frac{d\omega}{\omega} \frac{d t}{t}
\delta(1 - x - \omega t) \right]_+,
\eeq
where  $\omega \, \equiv \, 2E_g/m_Z$, is the normalized 
energy of the radiated gluon and $t \, \equiv \, 2(1-\cos\theta)\simeq
\theta^2$, 
with $\theta$ being the emission angle.
The $\delta$-function contains a typical kinematical constraint and
the `plus' regularization comes after including the virtual diagrams.
To factorize the kinematical constraint for multiple gluon emissions, a 
transformation to $N$-space is usually made:
\beq
\int_0^1 dx \, x^{N-1} 
\, \left[ \int_0^1 \, \int_0^1 \frac{d\omega}{\omega} \frac{d t}{t}
\delta(1 - x - \omega t) \right]_+
\, = 
\, \int_0^1 \int_0^1 
\frac{d\omega}{\omega} \frac{d t}{t}
\left[ \left(1 \, - \, \omega \, t \right)^{N-1} \, - \, 1 \right]
\, \simeq \, - \, \frac{1}{2} \, \ln^2 N.
\eeq
At higher orders, the large-$x$ logarithms are associated with multiple soft 
emissions ordered in angle:
\beq
\theta_1 \, > \, \theta_2 \, > \, \cdots \, > \, \theta_n.
\label{ang}
\eeq
Colour coherence \cite{dok}, in fact, dictates  that multiple soft radiation 
interferes destructively outside the
angular-ordered region (\ref{ang}) of the phase space.
That produces integrals in moment-space of the form
\begin{eqnarray}
\label{complicato}
~~~\alpha_S^n \!\!\!\!\!\!\!&&\!\!\!\!\!\!\!
\int_0^1 \int_0^1 \frac{d\omega_1}{\omega_1} \frac{d t_1}{t_1}
\int_0^1 \int_0^1 \frac{d\omega_2}{\omega_2} \frac{d t_2}{t_2}
\, \cdots \, 
\int_0^1 \int_0^1 \frac{d\omega_n}{\omega_n} \frac{d t_n}{ t_n } \,
\Theta \Big(t_1 \, > \, t_2 \, > \, \cdots \, > \, t_n \Big)
\nonumber\\
&&
\times \,
\left[ \left(1 \, - \, \omega_1 \, t_1 \right)^{N-1} \, - \, 1 \right]
\left[ \left(1 \, - \, \omega_2 \, t_2 \right)^{N-1} \, - \, 1 \right]
\, \cdots \, 
\left[ \left(1 \, - \, \omega_n \, t_n \right)^{N-1} \, - \, 1 \right]
\nonumber\\
&=& 
\frac{\alpha_S^n}{n!} 
\left( \int_0^1 \int_0^1 
\frac{d\omega}{\omega} \frac{d t}{t}
\left[ \left(1 \, - \, \omega \, t \right)^{N-1} \, - \, 1 \right]\right)^n ,
\end{eqnarray}
with 
$\omega_i \, \equiv \, 2E_i/m_Z$ and
and $t_i \, \equiv \, (1-\cos\theta_i)/2\,\simeq\,\theta_i^2$. 
\footnote{
The integral on the l.h.s. of Eq.~(\ref{complicato}) 
is completely factorized into 
single-particle integrals except for the angular-ordering $\Theta$-function.
This constraint can be eliminated by symmetrizing over the $t_i$'s:
\beq
\Theta \Big(t_1 \, > \, t_2 \, > \, \cdots \, > \, t_n \Big)
\, \to \,
\frac{1}{n!} \, \sum_{perm} 
\Theta \Big(t_{i1} \, > \, t_{i2} \, > \, \cdots \, > \, t_{in} \Big)
\, = \, \frac{1}{n!}.
\eeq
}
Eq.~(\ref{complicato}) implies that also the threshold logarithms have
an exponential structure. Schematically, the inclusion of higher orders
amounts to the replacement:
\beq
1 \, + \, \alpha_S \, \ln^2 N ~ \to ~ e^{ \alpha_S \, \ln^2 N}.
\eeq
A consistent method to accomplish the resummation of
mass (\ref{masslogs}) and threshold (\ref{xlogs}) logarithms is the
formalism of perturbative fragmentation
functions \cite{mele}. The basic idea is that 
a heavy quark is first produced
at large transverse momentum, $k_\perp\gg m$, and can be
treated in a massless fashion,
afterwards it slows down and fragments into a massive
parton. This leads to writing
the energy distribution of the $b$ quark, up to power corrections,
as the following convolution:
\beq
\frac{1}{\sigma} \, \frac{d\sigma}{dx}(x; \, m_Z^2, m_b^2)
\, = \, C(x; \, m_Z^2, \mu_F^2) \otimes D(x; \mu_F^2, m_b^2),
\label{con}
\eeq
where:
\begin{enumerate}
\item 
$C(x; \, m_Z^2, \mu_F^2)$ is a coefficient function,
obtained from a massless computation in a given factorization scheme, 
describing the emission off a light parton. 
The coefficient function contains large-$x$
logarithms, as in Eq.~(\ref{xlogs}), which are process-dependent;
\item
$D(x; \mu_F^2, m_b^2)$ is the perturbative fragmentation function,
associated with the transition of a massless parton into the heavy $b$. 
\end{enumerate}
The perturbative fragmentation function follows the 
Dokshitzer--Gribov--Lipatov--Altarelli--Parisi
(DGLAP) evolution equations \cite{dgl,ap}, which can be solved
once an initial condition is provided.
Solving the DGLAP equations one resums the large logarithms (\ref{masslogs})
which appear in the massive $b$-quark spectrum.
Function $D(x;\mu_F,m_b)$ can
therefore be factorized as:
\beq
D(x; \mu_F^2, m_b^2) \, = \, E(x; \mu_F^2, \mu_{0F}^2) \otimes 
D^{\rm ini}(x; \, \mu_{0F}^2, m_b^2), 
\eeq
where:
\begin{enumerate}
\item
$E(x; \mu_F^2, \mu_{0F}^2)$ is an evolution operator
from the scale $\mu_F$, typically of 
the order of the hard scale $m_Z$, down to 
$\mu_{0F}$, of the order of the bottom-quark
mass $m_b$. In $E(x;\mu_F^2,\mu_{0F}^2)$, the 
mass logarithms (\ref{masslogs}) are resummed;
\item
$D^{\rm ini}(x; \, \mu_{0F}^2, m_b^2)$ is the initial condition of the 
perturbative fragmentation function
at the scale $\mu_{0F}\simeq m_b$, first calculated in \cite{mele}, and
lately proved to be process-independent \cite{cc}.
It also contains threshold logarithms (\ref{xlogs}),
which are also process-independent, and were
resummed in \cite{cc} in the NLL approximation.
\end{enumerate}

The approach of perturbative fragmentation functions, with 
the inclusion of NLL large-$x$ resummation, has been applied to investigate 
$b$-quark production in $e^+e^-$ annihilation \cite{cc},
top ($t\to bW$) \cite{cm,ccm} and Higgs decays 
($H\to b\bar b$) \cite{corc}.
In this paper we shall go beyond the NLL approximation and, as far as the
perturbative calculation is concerned, we shall also 
include next-to-next-to leading logarithmic
(NNLL) threshold contributions 
to both $e^+e^-$ coefficient function and initial 
condition of the perturbative fragmentation function.
Also, the Mellin transforms of our resummed expressions will be performed
exactly, in order to resum constants and power-suppressed terms as well.

Most previous analyses convoluted the
parton-level spectrum with a non-perturbative fragmentation
function, which contains few parameters which are to be fitted to 
experimental data.
Afterwards, such models are used to predict the $B$ spectrum in other
processes, as long as one consistently describes perturbative $b$ 
production (see, e.g., Ref.~\cite{drol}).
An alternative method \cite{canas} 
consists in using data in Mellin moment space, 
such as the DELPHI ones \cite{delphi}, and fit directly the moments of the 
non-perturbative fragmentation function, without assuming any functional form 
for the hadronization model in $x$-space.

In this paper we reconsider the inclusion of non-perturbative
corrections to bottom-quark fragmentation. 
As discussed before,
instead of fitting the parameters of
a hadronization model, or the moments of the non-perturbative fragmentation
function, we model non-perturbative effects by the use of an 
analytic effective coupling constant
\cite{ermolaev,shirkov,stefanis}, which does not contain
any free parameter, extending the analysis carried out in
\cite{agl} and in \cite{noi4} in the framework of heavy-flavour decays.
Our modified coupling constant does not exhibit the Landau pole any longer,
and includes all-order absorptive
effects due to gluon branching.
We shall then be able to compare our predictions directly
with data, without using any extra hadronization model.

The plan of our paper is the following. 
In Section~\ref{sec2} we review the main features of the \msbar $e^+e^-$ 
coefficient function.
In Section~\ref{sec3} we discuss the massive computation,
the perturbative fragmentation approach and the resummation
of the large mass logarithms. 
In Sections~\ref{sec4} and \ref{sec5} 
we discuss NNLL large-$x$ resummation in the coefficient function and
initial condition of the perturbative fragmentation function, respectively,
pointing out the differences of our analysis with respect to 
previous ones.
In Section~\ref{sec6} we construct the analytic coupling constant
without the Landau pole.
In Section~\ref{sec7} we present our model,
based on an effective coupling constant 
as the only source of non-perturbative corrections to $b$-quark
fragmentation.
In Section~\ref{sec8} we present our results on the $B$-hadron spectrum 
in $e^+e^-$ annihilation 
in $x$-space, and investigate how they fare against SLD, ALEPH and OPAL data.
In Section~\ref{sec9} we perform a similar
analysis in $N$-space and compare with the DELPHI moments.
In Section~\ref{sec10} we summarize the main results of our work
and discuss the lines of development of our study.
There is also an appendix describing an algorithm to compute
numerically the Mellin transforms with the Fast Fourier 
Transform.

\sect{Massless-quark production and NLO coefficient function}
\label{sec2}

In this section we consider massless-quark production and
discuss the main features of the NLO
$e^+e^-$ coefficient function, which will be used later on
in the framework of
perturbative fragmentation functions.
The NLO computation exhibits many properties of the higher
orders, so that the general structure of the perturbative corrections 
can be understood by looking in detail into this case.

We study the production of a light-quark pair 
at the $Z^0$ pole, in the NLO approximation:
\begin{equation}
e^+ e^- \, \to \, Z^0(q) \, \to \, q(p_q) \, + \, 
\bar q (p_{\bar q}) \,  + \, (g(p_g)),
\end{equation}
where $(g(p_g))$ denotes a real (virtual) gluon, and
define the light-quark energy fraction as:
\begin{equation}
x\equiv {{2p_q\cdot q}\over {m_Z^2}}
\end{equation}
The differential cross section, in dimensional regularization, can be read
from the formulas in \cite{marti}:
\begin{equation}
\label{massless_cr}
{1\over \sigma} {{d\sigma_q}\over{dx}} =
\delta(1-x)
\, + \, \frac{\alpha_S(\mu_R^2)\, C_F}{\pi}
\left\{
\frac{1}{2} \left[ \frac{1+x^2}{1-x} \right]_+ \left(\ln\frac{m_Z^2}{\mu_F^2}
\, - \, {1\over{\hat\epsilon}}\right)+\hat A(x)\right\}+{\cal O}(\alpha_S^2), 
\end{equation}
where 
\beq
\frac{1}{\hat{\epsilon}} \, \equiv \, \frac{1}{\epsilon} \, - \, \gamma_E 
\, + \, \ln (4\pi),
\eeq
$\epsilon \, \equiv \, (4-D)/2$, $D$ is the number of space-time 
dimensions, $\mu_R$ and
$\mu_F$ are the renormalization and factorization scales, 
$\gamma_E \, = \, 0.577216 \dots$ is the Euler constant, $\alpha_S$ is the 
dimensionless strong coupling constant, $C_F=4/3$ and $\hat A(x)$ is 
the following function, independent
of $\hat\epsilon$ and $\mu_F$:
\beq
\hat A(x)=
 \left[\frac{\ln(1-x)}{1-x}\right]_+ 
- \frac{3}{4} \, \frac{1}{\left[ 1 - x\right]_+}
+\left( \frac{\pi^2}{3} - 3 \right) \, \delta(1-x) 
- \frac{1+x}{2} \, \ln(1-x)
+\frac{1+x^2}{1-x} \, \ln x
+ \frac{5-3x}{4}.
\eeq	
\beq
\sigma \,  = \, \sigma_0 
\left[
1 \, + \, {3\over 4}\frac{\alpha_S(\mu_R^2)C_F}
{\pi} \right]+{\cal O}(\alpha_S^2) 
\eeq
is the NLO total cross section, with $\sigma_0$ being the Born one.
Eq.~(\ref{massless_cr}) presents a pole, $\sim 1/\hat\epsilon$, which is 
remnant of the collinear singularity, and disappears in the total
cross section.
Subtracting the collinear
pole, one obtains what, in the
perturbative fragmentation formalism, is called the \msbar
coefficient function \cite{mele}:
\footnote{Unlike the usual convention, where the coefficient function is 
expressed in terms of $1/\sigma_0 (d\sigma/dx)$, 
we have divided by the NLO cross section
$\sigma$, so that the total integral is 1. 
In the following, we shall compare with data, also normalized to 1.} 
\begin{equation}
\label{cf}
\left[{1\over \sigma} {{d\hat\sigma_q}\over{dx}}\right]^{\overline{\rm MS}} =
\delta(1-x)
\, + \, \frac{\alpha_S(\mu_R^2)\, C_F}{\pi}
\left\{
\frac{1}{2} \left[ \frac{1+x^2}{1-x} \right]_+ \ln\frac{m_Z^2}{\mu_F^2}
\, + \, \hat A(x)\right\}+{\cal O}(\alpha_S^2).
\end{equation}
The coefficient function presents: 
\begin{enumerate}
\item
a term of collinear origin, 
\beq{{\alpha_S}\over{2\pi}}
P^{(0)}_{qq}(x)\ln{{m_Z^2}\over{\mu_F^2}}=
\frac{\alpha_S\, C_F}{2\pi} \, \left[ \frac{1+x^2}{1-x} \right]_+ 
\ln\frac{m_Z^2}{\mu_F^2},
\eeq
where $P^{(0)}_{qq}(x)$ is the leading-order
Altarelli--Parisi splitting function, 
containing contributions enhanced in the threshold region $x\to 1$;
\item
\label{enhance}
two terms which are enhanced at large $x$ and are independent of 
$\mu_F$:
\beq
\frac{\alpha_S\, C_F}{\pi} \, \left[\frac{\ln(1-x)}{1-x}\right]_+ 
~~~~~ {\mathrm{and}} ~~~~~
- \, \frac{3}{4} \,
\frac{\alpha_S\, C_F}{\pi} \, \frac{1}{~\left[ 1 - x \right]_+ };
\eeq
\item
a term proportional to $\delta(1-x)$, i.e. a spike in the elastic point 
$x\, = \, 1$:
\beq
\frac{\alpha_S\, C_F}{\pi} \,
\left( \frac{\pi^2}{3} - 3 \right) \, \delta(1-x);
\eeq
\item
terms dependent on $x$ and divergent at most logarithmically for
$x\to 1$:
\beq
\frac{\alpha_S\, C_F}{\pi} \,
\left[
\, - \, \frac{1+x}{2} \, \ln(1-x)
\, + \, \frac{1+x^2}{1-x} \, \ln x
\, + \, \frac{5 - 3 x}{4}
\right].
\eeq
We observe that the latter contribution also contains a term,
$\sim\ln x$, which is enhanced at small $x$. 
\end{enumerate}

We wish to rearrange the coefficient function and
put it into a form which will become more useful in the following sections,
when dealing with large-$x$ resummation.
To this goal, we rearrange the Altarelli--Parisi splitting function  
by using the identity:
\beq
\label{ident1}
\left[{1+x^2}\over{1-x}\right]_+
\, = \, \frac{2}{\, \left[ 1 - x \right]_+}
\, - \, (1+x)
\, + \, \frac{3}{2}\delta(1-x).
\eeq
Moreover, we factorize the constants which multiply the term
$\sim\delta(1-x)$, and
finally write the \msbar coefficient function in the form:
\begin{eqnarray}
\Bigg({1\over \sigma}\!\!\!\!\!\!\!\!&&\!\!\!\!\!\! {{d\hat{\sigma}_q}\over{dx}}\Bigg)^{\overline{\rm MS}}\!\!\!\!
=\Bigg[
1 +\frac{\alpha_S(\mu_R^2)C_F}{\pi} 
\Bigg( \frac{\pi^2}{3} - 3  + \frac{3}{4} \ln\frac{m_Z^2}{\mu_F^2} \Bigg)
\Bigg]\Bigg\{
\delta(1-x)+ \frac{\alpha_S(\mu_R^2)C_F}{\pi}  
\Bigg[\Bigg({1\over{1-x}}
\ln{{m_Z^2(1-x)}\over{\mu_F^2}}\Bigg)_+\!\! 
\nonumber\\
&-&\!\!\! {3\over {4(1-x)_+}}\Bigg]\Bigg\}
+ \frac{\alpha_S(\mu_R^2)C_F}{\pi}\left\{
{{1+x^2}\over{1-x}}\ln x
-\frac{1+x}{2} \, \left[ \ln(1-x) + \ln \frac{m_Z^2}{\mu_F^2} \right]
+ \frac{5-3x}{4}\right\}+{\cal O}(\alpha_S^2),
\label{cff}\nonumber\\
\end{eqnarray}
which is equal to (\ref{cff}) at ${\cal O}(\alpha_S)$.
Eq.~(\ref{cff}) exhibits the logarithm
$\ln[m^2_Z(1-x)/\mu_F^2]$, which
originates from the integration over the variable
$k^2=(p_q+p_g)^2(1-x)$:
\beq
\label{logc}
 C_F \, \frac{\alpha_S}{\pi} \, \ln \frac{m_Z^2(1-x)}{\mu_F^2} 
\, = \, \int_{\mu_F^2}^{m_Z^2(1-x)} 
\frac{dk^2}{k^2}  \, C_F \, \frac{\alpha_S}{\pi},
\eeq
where the constant term $C_F \, \alpha_S/\pi$ has been brought inside the
integral for reasons which will become clear later.
For soft and collinear radiation, $k^2\simeq E_g^2\theta_{qg}^2=
k_{\perp}^2$, the gluon transverse
momentum with respect to $q$.
In principle, the lower value of $k^2$ would be zero, as 
massless quarks can emit soft gluons at arbitrarily small angles. 
In dimensional regularization, using the
\msbar factorization scheme, the minimum $k^2$ is set by the
factorization scale:
$k^2_{\mathrm{min}}=\mu_F^2$. The upper limit in the integral (\ref{logc})
can be obtained observing that $k^2= (p_q+p_g)^2(1-x)=
(q-p_{\bar q})^2(1-x)\leq m_Z^2(1-x)$.

\sect{Heavy-quark production\\ and NLO perturbative
fragmentation function}
\label{sec3}

Let us now consider the production of massive bottom quarks 
in the NLO approximation:
\begin{equation}
e^+ e^- \, \to \, Z^0(q) \, \to \, b(p_b) \, + 
\, \bar b (p_{\bar b}) \,  + \, (g(p_g)).
\end{equation}
The differential cross section for the production of a $b$ quark of
energy fraction $x$ reads \cite{mele}:
\begin{eqnarray}\label{massive}
{1\over \sigma} {{d\sigma_b}\over{dx}} &=&
\delta(1-x)
\, + \, \frac{\alpha_S(\mu_R^2)\, C_F}{\pi}
\Bigg\{
\frac{1}{2} \left[ \frac{1+x^2}{1-x} \right]_+ \ln\frac{m_Z^2}{m_b^2}
\, - \, \left[\frac{\ln(1-x)}{1-x}\right]_+
\, - \, \frac{7}{4} \, \frac{1}{\left( 1 - x\right)_+}
\\
& + & \left(\frac{\pi^2}{3} - 2 \right) \, \delta(1-x)
\, + \, \frac{1+x}{2} \, \ln(1-x)
\, + \, \frac{1+x^2}{1-x} \, \ln x
\, + \, \frac{7-x}{4} 
\Bigg\} 
 + {\cal O}\left[\alpha_S(\mu_R^2) \left(\frac{m_b}{m_Z}\right)^p \right],
\nonumber
\end{eqnarray}
where $p\geq 1$.
The massive spectrum, unlike the massless one,
is free from collinear singularities because the quark mass acts a regulator.
However, Eq.~(\ref{massive}) presents a large mass logarithm,
$\sim\alpha_S\ln (m_Z^2/m_b^2)$, which needs to be resummed to improve the
perturbative prediction.

The resummation of $\ln(m_Z^2/m_b^2)$ can be achieved by the use of the
approach of heavy-quark perturbative fragmentation functions \cite{mele}, 
which factorizes 
the spectrum of a massive quark as the following convolution: 
\beq
\label{conv}
{1\over \sigma} {{d\sigma_b}\over{dx}} (x; \, m_Z^2,m_b^2)  \, = \,
\sum_i\int_{x}^1
{{{dz}\over z}\left[{1\over \sigma}
{{d\hat\sigma_i}\over {dz}}(z,m_Z^2,\mu_F^2)
\right]^{\overline{\mathrm{MS}}}
D_i^{\overline{\mathrm{MS}}}\left({x\over z}; \, \mu_F^2,m_b^2 \right)} \, + \,
{\cal O}\left[\left(\frac{m_b}{m_Z}\right)^p\right].
\eeq
In Eq.~(\ref{conv}),  
$1/\sigma\, (d\hat\sigma_i/dx)$ is the coefficient function,
corresponding to the production of a massless parton $i$, 
$D_i(x,m_b^2,\mu_F^2)$ is the heavy-quark perturbative fragmentation function,
associated with the transition of the massless parton $i$ into a heavy $b$,
and $\mu_F$ is the factorization scale. 
In the following, we shall neglect $b$ production via gluon splitting
$g\to b\bar b$, which is negligible at the $Z^0$ peak and suppressed
at large $x$.
Hence, in Eq.~(\ref{conv}), $i=b$ and $1/\sigma\, (d\hat\sigma_b/dz)$ is the
quark coefficient function, presented in Eq.~(\ref{cf}) in the \msbar
factorization scheme.
The perturbative fragmentation function
$D_b^{\overline{\mathrm{MS}}}$ expresses the fragmentation 
of a massless $b$ into a massive $b$.

Requiring the massive cross section to be independent of $\mu_F$, one
obtains that the perturbative fragmentation function follows the
DGLAP evolution equations \cite{dgl,ap}, 
which can be solved once an initial condition
is given. The initial condition of the perturbative
fragmentation function $D_b^{\rm ini}(x_b,\mu_{0F}^2,m_b^2)$ 
was given in \cite{mele}, and can be obtained
inserting in Eq.~(\ref{conv}) the massive
spectrum (\ref{massive}) and the \msbar
coefficient function (\ref{cf}).
It reads:
\begin{equation}\label{dbini}
D_b^{\mathrm{ini}}(x; \alpha_S(\mu_{0R}^2),\mu_{0R}^2,\mu_{0F}^2, m_b^2 )= 
\delta(1-x)+
{{\alpha_S(\mu_{0R}^2) C_F}\over{\pi}}
\!\left[ {{1+x^2}\over{1-x}} \left( \frac{1}{2} \ln {{\mu_{0F}^2}\over{m_b^2}}
- \ln (1-x)- \frac{1}{2}\! \right)\!\right]_+\!\!\!
+{\cal O}(\alpha_S^2).
\end{equation}
In \cite{cc}, 
the process-independence of the initial condition (\ref{dbini})
was established on general grounds. 
The solution of the DGLAP equations is typically obtained in Mellin moment
space. At NLO, and for an evolution from $\mu_{0F}$ to $\mu_F$, it is
given by:
\begin{equation}
\label{dbn}
D_{b,N}(\mu_F^2,m_b^2)\, = \,
E_N\left[ \alpha_S(\mu_{0F}^2), \alpha_S(\mu_F^2) \right]
D_{b,N}^{\rm ini}\left[ \alpha_S(\mu_{0R}^2), \mu_{0R}^2, \mu^2_{0F}, m^2_b 
\right],
\end{equation}
where $D_{b,N}^{\rm ini}$ is the Mellin transform of Eq.~(\ref{dbini})
and $E_N$ is the DGLAP evolution operator \cite{dgl,ap}:
\begin{equation}
E_N\left[ \alpha_S(\mu_{0F}^2), \alpha_S(\mu_F^2) \right] \, = \,
\exp\left\{ {{P_N^{(0)}}\over{2\pi \beta_0}}
\ln{{\alpha_S(\mu^2_{0F})}\over {\alpha_S(\mu^2_F)}}
+{{\alpha_S(\mu^2_{0F})-\alpha_S(\mu^2_F)}\over{4\pi^2\beta_0}}
\left[P_N^{(1)}-{{2\pi \beta_1}\over {\beta_0}}P_N^{(0)}\right]\right\}.
\label{evol}
\end{equation}
In Eq.~(\ref{evol}), 
$P_N^{(0)}$ and $P_N^{(1)}$ are the Mellin transforms of the LO and
NLO splitting functions;
$\beta_0$ and $\beta_1$ are the first two coefficients of the QCD 
$\beta$-function:
\begin{equation}
\beta_0 \, = \, {{33 - 2n_f}\over {12\pi}},\ \ 
\beta_1 \, = \, {{153 - 19n_f}\over{24\pi^2}},
\label{b0b1}
\end{equation}
which enter in the NLO expression of the strong coupling constant:
\begin{equation}
\alpha_S(Q^2) \, = \, {1\over {\beta_0\ln(Q^2/\Lambda^2)}}
\left\{ 1-{{\beta_1\ln\left[\ln (Q^2/\Lambda^2)\right]}\over
{\beta_0^2\ln(Q^2/\Lambda^2)}}\right\}.
\label{alphanlo}
\end{equation}
As discussed in \cite{mele}, Eq.~(\ref{dbn}) resums leading (LL)
$\alpha_S^n(\mu_F^2)\ln^n(\mu_F^2/\mu_{0F}^2)$ and next-to-leading (NLL) 
$\alpha_S^n(\mu_F^2) \ln^{n-1}(\mu_F^2/\mu_{0F}^2)$ logarithms 
of the ratio of the two factorization scales (collinear
resummation).
For $\mu_F\simeq m_Z$ and $\mu_{0F}\simeq m_b$, as we shall
assume hereafter, one resums the large logarithm $\ln(m_Z^2/m_b^2)$,
which appears in the NLO massive spectrum (\ref{massive}), with NLL accuracy.

As we did for the coefficient function, we rearrange the initial
condition of the perturbative fragmentation function into a form which
will be convenient when discussing soft-gluon resummation.
We use the identity (\ref{ident1}) and the relation
\beq
\left[ \frac{1+x^2}{1-x} \, \ln (1-x) \right]_+
\, = \, 
2 \left[ \frac{\ln(1-x)}{1-x} \right]_+
\, - \, (1+x)\ln(1-x) 
\, - \, \frac{7}{4} \delta(1-x).
\eeq
Furthermore, we factorize the coefficient of the term 
$\sim\delta(1-x)$ and write $D_b^{\mathrm{ini}}$ in the
following form, which is equivalent to (\ref{dbini}), up to
terms of ${\cal O}(\alpha_S^2)$:
\begin{eqnarray}
\label{heuristic2}
D_b^{\mathrm{ini}}[x;\alpha_S(\mu_{0R}^2), \mu_{0F}^2, m_b^2 ] & = & 
\left[1+{{\alpha_S(\mu_{0R}^2)  C_F}\over{\pi}}
\left( 1+\frac{3}{4} \ln {{\mu_{0F}^2}\over{m_b^2}}  \right)
\right]
\Bigg\{\delta(1-x)\nonumber\\
&+&\left. {{\alpha_S(\mu_{0R}^2) C_F}\over{\pi}} \left[ 
\left({1\over{1-x}} 
\ln{{\mu_{0F}^2}\over{m_b^2(1-x)^2}} \right)_+ 
- {1\over{(1-x)_+}}\right]\right\}\nonumber\\
&+&{{\alpha_S(\mu_{0R}^2)  C_F}\over{\pi}}
(1+x) \left[ \frac{1}{2} 
- \frac{1}{2} \ln {{\mu_{0F}^2}\over{m_b^2}} + \ln(1-x) \right]
+{\cal O}(\alpha_S^2).
\end{eqnarray}
The logarithm $\ln[\mu_{0F}^2/(m_b^2(1-x)^2)]$
comes again from an integral over $k^2=(p_b+p_g)^2(1-x)$:
\beq
\label{logd}
 C_F \, \frac{\alpha_S}{\pi} \,
\ln {{\mu_{0F}^2}\over{m_b^2(1-x)^2}}
\, = \, \int^{\mu_{0F}^2}_{m_b^2(1-x)^2} \frac{dk^2}{k^2}
 \, C_F \, \frac{\alpha_S}{\pi}.
\eeq
As in (\ref{logc}), for soft and small-angle radiation
$k^2\simeq 
k_\perp^2$, the transverse momentum of the gluon relative to the $b$.
Following \cite{cc}, the lower limit of the $k^2$-integration is easily found 
by considering the dead-cone effect \cite{dok}.
In fact, unlike the coefficient function, where quarks are treated as 
massless, soft radiation off a massive $b$ quark 
is suppressed at angles lower than
\footnote{The bremsstrahlung spectrum off a heavy quark
reads: $d\sigma\sim \alpha_S\  (d\omega/\omega)\ 
 [d\theta^2/(\theta^2+\theta_{\rm min}^2)]$, where $\omega$ is the
energy of the radiated soft gluon.}:
\beq\label{dead}
\theta_{\min} \, \simeq \, \frac{m_b}{E_b}.
\eeq
To the logarithmic accuracy we are interested in,
we can neglect emission inside the dead cone and 
use Eq.~(\ref{dead}) to obtain the lower limit on the
emitted-gluon transverse momentum:
\beq
k^2_{\min}\simeq 
k^2_{\perp \min} \, \simeq \, E_g^2 \, \theta^2_{\min} \, \simeq \, m_b^2
(1-x)^2 .
\eeq
Comparing Eq.~(\ref{logd}) with (\ref{logc}), we observe that, in the
\msbar scheme, the factorization scale squared $\mu^2_F$ is
the lower limit on the gluon transverse momentum 
in the coefficient function; $\mu^2_{0F}$ is instead
the upper limit on $k^2$ in the initial condition.
The operator $E(\alpha_S(\mu_{0F}^2),\alpha_S(\mu_F^2))$, given in 
Eq.~(\ref{evol}) describes the evolution between these two scales. 

Before closing this section, we
point out that, although we shall use NLO coefficient function and
initial condition, and evolve the perturbative fragmentation
function with NLL accuracy, in principle we could go beyond such
approximations. In fact, the NNLO $e^+e^-$ coefficient
function was calculated in \cite{neerven,alex4} and
NNLO corrections to the initial condition of the perturbative 
fragmentation function in \cite{alex1,alex2}. 
Furthermore, NNLO contributions to the non-singlet
time-like splitting functions were computed in \cite{alex3}, while
in the singlet sector they are still missing.
In any case, we shall delay the inclusion of such NNLO corrections to
future work.

\sect{Large-$x$ resummation in the coefficient function}
\label{sec4}

In this section we perform threshold resummation in the 
$e^+e^-$ coefficient function to next-to-next-to-leading
logarithmic accuracy, and  combine the resummed result 
with the first-order one presented in Section~\ref{sec2}.

\subsection{Resummed coefficient function in Mellin space}
\label{sec41}

According to Eq.~(\ref{cf}), the NLO $e^+e^-$ coefficient function 
contain terms of the form
\beq
\label{primi2}
\alpha_S \, \left[ \frac{\ln (1-x)}{1-x} \right]_+
~~~~~{\rm {and}}~~~~~
\alpha_S \, \frac{1}{ ~ \left[ 1-x \right]_+},
\eeq
which become large in the limit $x \, \to \, 1$, corresponding to soft
or collinear gluon radiation.
All-order resummation in the coefficient function can be performed following 
the general lines of Refs.~\cite{sterman,ct}, and was implemented 
in \cite{cc} in the NLL approximation.

Threshold resummation is typically performed 
in $N$-space \cite{ct,cc}, where kinematical constraints factorize and the
$x\to 1$ limit corresponds to $N\to\infty$. 
The Mellin transform of the NLO coefficient function,
which we denote by $C_N$, can be found in
\cite{mele}. At large $N$, it exhibits
single and double logarithms of the Mellin variable:
\begin{eqnarray}
C_N\left[ \alpha_S(\mu_R^2), \mu_R^2, \mu_F^2, m_Z^2 \right] &=& 1 \, + \,
{{\alpha_S(\mu_R^2)C_F}\over{\pi}}\left[{1\over 2}\ln^2 N+
\left( {3\over 4} + \gamma_E - \ln{{m_Z^2}\over{\mu_F^2}} \right)\ln N
\right.
\nonumber\\
&+& 
\, \left. Q(\mu_F^2,m_Z^2) \, + \, {\cal O}\left({1\over N}\right)\right],
\label{cflargen}
\end{eqnarray}
with the constant terms given by:
\begin{equation}
Q(\mu^2_F,m_Z^2) \, \equiv \, 
\left({3\over 4} - \gamma_E \right)\ln{{m_Z^2}\over{\mu_F^2}}
\, + \, {5\over{12}}\pi^2 - 3 + {1\over 2}\gamma_E^2+{3\over 4} \gamma_E.
\label{qq}
\end{equation}
Besides the contributions (\ref{primi2}), the
NLO coefficient function (\ref{cf}) also contains a contribution
$\sim\ln(1-x)$. However, its Mellin transform
behaves like $\sim\ln N/N$ and is therefore ${\cal O}(1/N)$ at large $N$.
The other terms in Eq.~(\ref{cf}) are suppressed at large $N$.

The resummed coefficient function has the following
generalized exponential structure:
\begin{equation}
\Delta_N^{(C)}\left[\alpha_S(\mu_R^2), \mu_R^2, \mu_F^2, m_Z^2\right] \, = \, 
\exp
\left\{
G_N^{(C)}\left[ \alpha_S(\mu_R^2), \mu_R^2, \mu_F^2, m_Z^2 \right]
\right\},
\end{equation}
where \cite{cc}
\beq
\label{GNC}
G_N^{(C)}\left[ \alpha_S(\mu_R^2), \mu_R^2, \mu_F^2, m_Z^2 \right] \, = \, 
\int_0^1 {dz {{z^{N-1}-1}\over{1-z}}}
\left\{\int_{\mu_F^2}^{m_Z^2 (1-z)} {{dk^2}\over {k^2}}
A\left[\alpha_S(k^2)\right] +
B\left[\alpha_S\left(m_Z^2(1-z)\right)\right]\right\}.
\eeq
The exponent
$G^{(C)}_N\left[ \alpha_S(\mu_R^2), \mu_R^2, \mu_{F}^2, m_Z^2 \right]$ 
resums the large logarithms appearing in Eq.~(\ref{cflargen}):
\beq
{\rm LL}: ~ \alpha_S^n \, \ln^{n+1}N;
~~~~~~~~
{\rm NLL}: ~ \alpha_S^n \, \ln^n N;
~~~~~~~~
{\rm NNLL}: ~ \alpha_S^n \, \ln^{n-1}N.
\eeq
As in \cite{ct}, the integration variables are 
$z \, = \, 1 - x_g$, $x_g$ being the gluon energy fraction, and 
$k^2 \, = \, (p_b+p_g)^2(1-z)$. In soft approximation, $z\simeq x$; for
small-angle radiation $k^2 \, \simeq \, k_{\perp}^2$, 
the gluon transverse momentum with respect to the $b$.

The functions $A(\alpha_S)$ and $B(\alpha_S)$ 
can be expanded as a series in $\alpha_S$ as:
\begin{equation}
A(\alpha_S) \, = \, \sum_{n=1}^{\infty}\left({{\alpha_S}\over
{\pi}}\right)^n A^{(n)}, 
\end{equation}
\begin{equation}
B(\alpha_S) \, = \, \sum_{n=1}^{\infty}\left({{\alpha_S}\over
{\pi}}\right)^n B^{(n)}.
\end{equation}
In the NLL approximation, one needs to include the first two coefficients
of $A(\alpha_S)$ and the first of $B(\alpha_S)$; to NNLL accuracy, 
$A^{(3)}$ and $B^{(2)}$ are also needed.
Their expressions read:
\begin{eqnarray}
A^{(1)}&=&C_F,\\
A^{(2)}&=& {1\over 2} C_F \left[ C_A\left(
{{67}\over{18}}-\zeta(2)\right) -{5\over 9}n_f\right],\\
A^{(3)}&=&C_F\left\{C_A^2\left[{{245}\over{96}}+{{11}\over{24}}\zeta(3)
-{{67}\over{36}}\zeta(2)+{{11}\over 8}\zeta(4)\right]-C_An_f
\left[{{209}\over{432}}+{7\over{12}}\zeta(3)-{5\over{18}}
\zeta(2)\right]\right.
\nonumber\\
&-&\left.C_Fn_f\left[{{55}\over{96}}-{{\zeta(3)}\over 2}\right]-
{{n_f^2}\over{108}}\right\},\\
B^{(1)}&=&-{3\over 4}C_F,\\
B^{(2)}&=&C_F\left[C_A\left(-{{3155}\over {864}}+{{11}\over{12}}\zeta(2)+
{5\over 2}\zeta(3)\right)-C_F\left({3\over{32}}+{3\over 2}\zeta(3)-
{3\over 4}\zeta(2)\right)\right.\nonumber\\
&+& \left.n_f\left({{247}\over{432}}-{{\zeta(2)}\over 6}\right)\right],
\end{eqnarray}
where $C_A \, = \, 3$, $n_f$ is the number of active flavours and 
$\zeta(x) 
\, \equiv \, \sum_{n=1}^{\infty} 1/n^x$ is the Riemann Zeta function. 
The first two coefficients of function $A(\alpha_S)$ have been known
for long time \cite{ct}; more recent is the calculation of $A^{(3)}$
\cite{moch1}.
Function $B(\alpha_S)$ is associated with the radiation off the unobserved
massless parton, e.g. the $\bar b$ if one detects the $b$: the coefficient
$B^{(1)}$ was given in \cite{ct}, $B^{(2)}$ was computed in \cite{moch2}.

We can already observe that the
integral over the transverse momentum in Eq.~(\ref{GNC})
is a generalization of the NLO integral 
in Eq.~(\ref{logc}), where 
\beq
C_F \, \frac{\alpha_S}{\pi} \, = \, A^{(1)} \frac{\alpha_S}{\pi}
\,\, \to \,\,  A^{(1)}{{\alpha_S(k^2)}\over{\pi}}  \,\, \to \,\,  
A\left[ \alpha_S(k^2) \right].
\eeq
Analogously, function $B(\alpha_S)$ generalizes the
coefficient of the $\sim 1/(1-x)_+$ term in (\ref{cff}):
\beq
- \, \frac{3}{4}  \, C_F \, \frac{\alpha_S}{\pi} 
\, = \, B^{(1)} \frac{\alpha_S}{\pi}
 \, \, \to \, \, 
B\left[ \alpha_S\left( m_Z^2(1-x) \right) \right].
\eeq
A delicate point  of our approach concerns the integral over $z$
in Eq.~(\ref{GNC}). 
Calculations which use the standard coupling constant $\alpha_S(k^2)$,
such as \cite{cc}, typically perform the $z$-integration approximating the 
term $\sim (z^{N-1}-1)$ in such a way that only logarithmically-enhanced
contributions $\sim\ln^k N$ are kept in the exponent. 
To NLL accuracy, such an approximation reads \cite{ct}:
\begin{equation}
z^{N-1}-1\to  \Theta\left( 1 - z - {{e^{-\gamma_E}}\over N}\right),
\label{step}
\end{equation}
where $\Theta(x)$ is the Heaviside function.
Beyond NLL, the prescription (\ref{step}) can be generalized as 
discussed in \cite{defl}. 
In fact, the integrations in $z$ and $k^2$ in (\ref{GNC}) involve
the infrared region $z\to 1$ for any value of $N$. As observed in
\cite{korste}, performing such integrations exactly, when 
using truncated expressions for $A(\alpha_S)$ and $B(\alpha_S)$, will lead 
to a factorial divergence, corresponding to a power correction which,
in our case, would be $\sim\Lambda/m_Z$. Ref.~\cite{braun}, however,
remarkably pointed out that this contribution is spurious as it is
actually related to the fact that one
employed truncated expressions for $A(\alpha_S)$ and $B(\alpha_S)$.
It was shown in \cite{braun} that the higher-order coefficients of
such functions also present factorial singularities leading to 
contributions $\sim \Lambda/m_Z$, which
cancel the analogous terms coming from the exact integration, so that 
one is just left with a power correction $\sim \Lambda^2/m_Z^2$. 
It was hence argued on general grounds in \cite{min,defl} 
that, when considering truncated
expressions for functions $A(\alpha_S)$ and $B(\alpha_S)$, employing the
step function  (\ref{step}) or its extensions beyond NLL \cite{defl}
is a better approximation than performing the $z$-integration
exactly, since the transformation (\ref{step}) and its generalizations
do not lead to any spurious factorial growth of the
cross section.

However, all such results are valid as long as one uses the 
standard coupling constant in the resummed formulas. 
As we shall detail in Sections \ref{sec6} and \ref{sec7}, 
we will model non-perturbative
corrections by means of an effective QCD coupling constant, 
based on an extension of the work in \cite{ermolaev, shirkov,stefanis}, which 
does not present the Landau pole any longer and includes power-suppressed
contributions.
As a result, some power corrections will be 
unavoidably transferred to the resummed coefficient function
when employing the
effective coupling in expressions like Eq.~(\ref{GNC}), independently of
how one performs the longitudinal-momentum integration.
To our knowledge,  there is currently no 
analysis, such as the ones in \cite{braun,min}, on 
factorial divergences and power corrections in the
coefficient function (\ref{GNC}), when using an effective 
analytic coupling constant along with 
truncated $A(\alpha_S)$ and $B(\alpha_S)$. We cannot therefore draw
any firm conclusion on whether, within our model, 
it is better to perform the
$z$-integration in an exact or approximated way. 
Given the phenomenological aim of the present paper,
we prefer to postpone such an investigation. In the following,
as already done in \cite{noi4} when the effective
coupling was used in the context of heavy-flavour
decays, we shall present results obtained performing the Mellin 
transform (\ref{GNC}) exactly. In other words, doing the $z$-integration 
in (\ref{GNC}) exactly should be seen, for the time being, as part of the
non-perturbative model which we shall propose.

Due to its complexity, we cannot express the result of the exact
longitudinal-momentum integration (\ref{GNC}) in a closed analytic form, 
but we shall perform it numerically.
The integration over $k^2$ 
could in principle be made analytically, and the result
expressed in terms of polylogarithms. For simplicity's sake, however,
we perform also the $k^2$-integration numerically.

For $0 \, \leq  \, z \, \leq 1$, the argument $k^2$ of
$\alpha_S(k^2)$ in Eq.~(\ref{GNC}) varies from zero to the hard scale 
squared $m^2_Z$;
this implies that the number of active quark 
flavours does change in the $k^2$-integration.
In the following, we shall include correctly 
the variation of $n_f$
at the quark-mass thresholds when doing the numerical integration.

Let us now see with an explicit analytic computation the difference
between an exact Mellin transform and an approximation using
the step function.
The expansion of the exponent of the resummed coefficient
function reads, to ${\cal O}(\alpha_S)$:
\begin{eqnarray}
\left(
\Delta_N^{(C)}
\left[ \alpha_S(\mu_R^2), \mu_R^2, \mu_F^2, m_Z^2 \right]\right)_
{\alpha_S}
&=&1+{{\alpha_S(\mu_R^2)}\over{\pi}} 
\int_0^1{dz} z^{N-1}\left\{A^{(1)}\left[{1\over{1-z}}
\ln{{m_Z^2(1-z)}\over{\mu_F^2}}\right]_+\right.\nonumber\\
&+&\left. B^{(1)}
\left[{1\over{1-z}}\right]_+\right\}\label{integr}\\
&=&1+ {{\alpha_S(\mu_R^2)C_F}\over{\pi}} 
\left\{{1\over 2} \left[ S_1^2(N-1) + S_2(N-1) \right]\right. \nonumber\\
&+& \left.\left({3\over 4}
-\ln{{m_Z^2}\over{\mu_F^2}}\right)S_1(N-1)\right\}.
\label{rescnlo}\end{eqnarray}
In Eq.~(\ref{rescnlo}) 
we have defined the harmonic sums $S_1(N)$ and $S_2(N)$, which are given by
\begin{eqnarray}
S_1(N) & \equiv & \psi_0(N+1) \, - \, \psi_0(1),
\\
S_2(N) & \equiv & -\psi_1(N+1) \, + \, \psi_1(1),
\end{eqnarray}
where
\begin{equation}
\psi_k(x) \, \equiv \, {{d^{k+1}\ln\Gamma(x)}\over {dx^{k+1}}}
\end{equation} 
are the polygamma function and
$\Gamma(x)$ is the Euler Gamma function. 
Using the large-$N$ expansions:
\begin{eqnarray}
S_1(N) & = & \ln N \, + \, \gamma_E \, + \, {\cal O}\left({1\over N}\right),
\\
S_2(N) & = & {{\pi^2}\over 6} \, + \, {\cal O}\left({1\over N}\right),
\end{eqnarray}
we can write for large $N$:
\begin{eqnarray}
\label{resclargen}
\left(\Delta^{(C)}_N\left[ \alpha_S(\mu_R^2), \mu_R^2, \mu_F^2, m_Z^2 
\right]\right)_{\alpha_S}
&=&1+  {{\alpha_S(\mu_R^2)C_F}\over{\pi}} 
\left[{1\over 2}\ln^2 N +
\left({3\over 4}+ \gamma_E - \ln{{m_Z^2}\over{\mu_F^2}} 
\right)\ln N\right.\nonumber\\
&+&\left.  Q'(\mu_F^2,m_Z^2) +{\cal O}\left({1\over N}\right)\right],
\end{eqnarray}
where $Q'(\mu_F^2,m_Z^2)$ collects the constant terms 
\begin{equation}
Q'(\mu^2_F,m_Z^2) \, \equiv \, 
{{\pi^2}\over{12}} \, + \, {{\gamma_E^2}\over 2} \, + \, {3\over 4}\gamma_E
\, - \, \gamma_E \, \ln{{m_Z^2}\over{\mu_F^2}}.
\label{q1}
\end{equation}
By comparing Eq.~(\ref{rescnlo}) with Eq.~(\ref{resclargen}), we explicitly 
see that performing the Mellin transform exactly, rather than with the
step-function approximation, 
amounts to including also constants and contributions
power-suppressed at large $N$ in the exponent $G_N^{(C)}$. 
If one used the $\Theta$-function approximation, as in \cite{cc}, the
${\cal O}(\alpha_S)$ expansion of the resummed coefficient function
would contain
only the logarithmically-enhanced contributions 
$\sim\alpha_S\ln N$ and $\sim\alpha_S\ln^2 N$.
The resummed coefficient function in the original $x$-space 
is finally obtained by an inverse Mellin transform:
\beq\label{inv}
\Delta^{(C)}\left[ x; \, \alpha_S(\mu_R^2), \mu_R^2, \mu_F^2, m_Z^2 \right] 
\, = \, \int_{c-i\infty}^{c+i\infty}
\frac{dN}{2\pi i}
\, x^{-N} \,
\Delta^{(C)}_N\left[ \alpha_S(\mu_R^2), \mu_R^2, \mu_F^2, m_Z^2 \right],
\eeq
where the (real) constant $c$ is chosen so that all the singularities
of $\Delta^{(C)}_N$ lie to the left of the integration contour.
The inverse transform (\ref{inv}) is also made exactly, in a numerical way.

\subsection{Matching of resummed and NLO coefficient function}
\label{sec42}

We wish to implement the matching of the resummed coefficient function
with the exact ${\cal O}(\alpha_S)$ one, 
in order to obtain a good approximation in the whole 
kinematical range. 
We can perform the matching in $N$-space and then invert the final 
result to $x$-space; alternatively, we can invert the resummed coefficient
function as in (\ref{inv}),
and then match it to the NLO $x$-space result.
Given the high accuracy of our approach and the delicate issue of the
exact Mellin transform, we have matched resummed and NLO 
coefficient function in both
$N$- and $x$-spaces, and checked the consistency of our results. We discuss
the matching in both spaces.
\subsubsection{$N$-space}
We would like to write the NNLL-resummed coefficient function as:
\begin{eqnarray}
\label{cfin}
C_N^{\rm res}\left[\alpha_S(\mu_R^2), \mu_R^2, \mu_F^2, m_Z^2 \right]  
&=& K^{(C)}\left[ \alpha_S(\mu_R^2), \mu_R^2, \mu_F^2, m_Z^2 \right] 
\Delta_N^{(C)}\left[ \alpha_S(\mu_R^2), \mu_R^2, \mu_F^2, m_Z^2 \right] 
\nonumber\\
&+& d_N^{(C)}\left[ \alpha_S(\mu_R^2), \mu_R^2, \mu_F^2, m_Z^2 \right],
\end{eqnarray}
where:
\begin{enumerate}
\item
$K^{(C)}\left[ \alpha_S(\mu_R^2), \mu_R^2, \mu_F^2, m_Z^2 \right]$
is a hard factor, introduced for the sake of including subleading
terms, which corresponds to the difference between the constant terms 
which are present in the exact NLO coefficient
function and the ones contained in the ${\cal O}(\alpha_S)$
expansion of the resummed result;
\item 
$\Delta^{(C)}_N\left[x; \, \alpha_S(\mu_R^2), \mu_R^2, \mu_F^2, m_Z^2 \right]$
is the resummed coefficient function, presented in (\ref{GNC}); 
\item
$d^{(C)}_N\left[\alpha_S(\mu_R^2), \mu_R^2, \mu_F^2, m_Z^2 \right]$
is a remainder function, 
collecting the left-over NLO contributions, which are suppressed at large $N$.
\end{enumerate}
The hard factor reads:
\begin{equation}\label{cn}
 K^{(C)}\left[ \alpha_S(\mu_R^2), \mu_R^2, \mu_F^2, m_Z^2 \right] =
1 + {{\alpha_S(\mu_R^2)\, C_F}\over{\pi}} \, Q'' (\mu_F^2, m_Z^2) ,
\end{equation}
with
\begin{equation}
Q''(\mu_F^2,m_Z^2) \, \equiv \, Q(\mu_F^2,m_Z^2) \, - \, Q'(\mu_F^2,m_Z^2)=
{3\over 4}\ln{{m_Z^2}\over{\mu_F^2}} \, + \, {{\pi^2}\over 3} \, - \, 3,
\end{equation}
where $Q(\mu_F^2,m_Z^2)$ and $Q'(\mu_F^2,m_Z^2)$ have been defined
in Eqs.~(\ref{qq}) and (\ref{q1}), respectively.
If we had used the step-function approximation, no constants would have been
resummed, and the hard factor would have contained all the constant terms
present in the NLO coefficient function, i.e. $Q(\mu_F^2,m_Z^2)$.

In $N$-space, the remainder function $d_N^{(C)}$ 
is obtained subtracting from the 
exact NLO coefficient function in Mellin space \cite{mele} the
${\cal O}(\alpha_S)$ expansion of $K^{(C)}\Delta^{(C)}$. 
We obtain:
\begin{eqnarray}
d_N^{(C)}\left[ \alpha_S(\mu_R^2), \mu_R^2, \mu_F^2, m_Z^2 \right]
&=&C_N\left[ \alpha_S(\mu_R^2), \mu_R^2, \mu_F^2, m_Z^2 \right]\nonumber\\
&-&
\left( K^{(C)}\left[ \alpha_S(\mu_R^2), \mu_R^2, \mu_F^2, m_Z^2 \right]
\Delta_N^{(C)}\left[ \alpha_S(\mu_R^2), \mu_R^2, \mu_F^2, m_Z^2 \right]
\right)_{\alpha_S}\nonumber\\
&=&
\frac{\alpha_S(\mu_R^2)}{ \pi } \, C_F
\left\{ -\frac{1}{2}
   \ln \frac{m_Z^2}{\mu_F^2}
   \left(\frac{1}{N+1} + \frac{1}{N}\right)+\frac{1}
   {2} \psi ^{(0)}(N)
   \left(\frac{1}{N+1} +\frac{1}{N}\right)\right.
\nonumber\\
&-&\left. 2 \psi^{(1)}(N) 
+  \frac{\gamma_E}{2} \left( \frac{1}{N} + \frac{1}{N+1} \right)
 + \frac{7}{4 N} 
- \frac{5}{4 (N+1)}\right.\nonumber\\
 &+&\left. \frac{3}{2}  \left[ \frac{1}{N^2} + \frac{1}{(N+1)^2} \right]
\right\}.\label{dnc}
\end{eqnarray}

\subsubsection{$x$-space}
Likewise, we wish to write the NNLL-resummed
coefficient function in $x$-space in a form
analogous to Eq.~(\ref{cfin}):
\begin{eqnarray}
\label{Ccompleto}
C^{\rm res}\left[ x; \, \alpha_S(\mu_R^2), \mu_R^2, \mu_F^2, m_Z^2 \right]  
&=& K^{(C)}\left[ \alpha_S(\mu_R^2), \mu_R^2, \mu_F^2, m_Z^2 \right] 
\, \Delta^{(C)}\left[ x; \, \alpha_S(\mu_R^2), \mu_R^2, \mu_F^2, m_Z^2 \right]
\nonumber\\
&+&  
d^{(C)}\left[ x; \, \alpha_S(\mu_R^2), \mu_R^2, \mu_F^2, m_Z^2 \right].
\end{eqnarray}
To get the remainder function and the constants, 
the factorized expression (\ref{cff}) of
the coefficient function turns out to be particularly useful.
First, we need the 
${\cal O}(\alpha_S)$ expansion of the resummed result in $x$-space, 
that can be read from the integrand function in the Mellin transform
(\ref{integr}):
\begin{equation}\label{asx}
\left(\Delta^{(C)}\left[ x;  \alpha_S(\mu_R^2), \mu_R^2, \mu_F^2, m_Z^2 
\right]\right)_{\alpha_S}=
\delta(1-x) +{{\alpha_S(\mu_R^2)C_F}\over{\pi}} \left\{
\left[\frac{1}{1-x} \ln \frac{m_Z^2 (1-x)}{\mu_F^2}
\right]_+ -{3\over{4[1-x]_+}}\right\}.
\end{equation}
Then, by 
comparing Eq.~(\ref{asx}) with Eq.~(\ref{cff}), 
we obtain the constants $K^{(C)}$, i.e. the coefficient of the term 
$\sim\alpha_S\, \delta(1-x)$ in Eq.~(\ref{cff}), 
obviously equal to the ones in Eq.~(\ref{cn}), and the
remainder function:
\begin{eqnarray}
\label{dxc}
d^{(C)}\left[ x; \, \alpha_S(\mu_R^2), \, \mu_R^2, \mu_F^2, m_Z^2 \right] 
&=&C\left[ x;  \alpha_S(\mu_R^2), \mu_R^2, \mu_F^2, m_Z^2 
\right]\nonumber\\
&-&\left( K^{(C)}\left[ \alpha_S(\mu_R^2), \mu_R^2, \mu_F^2, m_Z^2 \right]
\Delta^{(C)}
\left[ x;  \alpha_S(\mu_R^2), \mu_R^2, \mu_F^2, m_Z^2 
\right]\right)_{\alpha_S}\nonumber\\
&=& 
\frac{\alpha_S(\mu_R^2) }{\pi } \, C_F
\Bigg\{
\frac{1}{4} ( 5 - 3 x)
\, - \, \frac{1}{2} (1+x) \, \left[ \ln (1-x) \, + 
\, \ln \frac{m_Z^2}{\mu_F^2} \right]
\nonumber\\
&+& \frac{1+x^2}{1-x} \ln x
\Bigg\}.
\end{eqnarray}
In (\ref{dxc}), we denoted by 
$C\left[ x;  \alpha_S(\mu_R^2), \mu_R^2, \mu_F^2, m_Z^2 \right]$
the NLO \msbar coefficient function (\ref{cf}). 
Of course, the Mellin transform of Eq.~(\ref{dxc}) yields Eq.~(\ref{dnc}).

\sect{Large-$x$ resummation in the initial condition}
\label{sec5}

In this section we resum in the NNLL approximation 
the threshold logarithms which
appear in the initial condition of the perturbative
fragmentation function, and we match the
resummed expression with the exact NLO one.
As discussed in \cite{cc}, large-$x$ resummation in the initial condition
is process-independent.

\subsection{Resummed initial condition in Mellin space}
The initial condition (\ref{dbini}) present terms $\sim[\ln(1-x)/(1-x)]_+$
and $1/(1-x)_+$ which are to be resummed to all orders. 
In $N$-space, the Mellin transform of the NLO initial condition in 
Eq.~(\ref{dbini}) exhibits
single and double logarithms of $N$:
\begin{eqnarray}
D_N^{\rm ini}(\alpha_S(\mu_{0R}^2), \mu_{0R}^2, \mu_{0F}^2, m_b^2) &=& 
1 \, + \,
{{\alpha_S(\mu_{0R}^2)\, C_F}\over{\pi}} \left[ - \ln^2 N+
\left(\ln{{m_b^2}\over{\mu_{0F}^2}} - 2 \gamma_E + 1 \right)\ln N\right.
\nonumber\\
&+& \left.
Y(\mu_{0F}^2, m_b^2) \, + \, {\cal O}\left({1\over N}\right)
\right],
\label{dnlolargen}
\end{eqnarray}
where the constant terms are given by:
\begin{equation}
Y(\mu_{0F}^2, m_b^2)=
1 - 
{{\pi^2}\over 6} + \gamma_E - \gamma_E^2 + \left(\gamma_E - {3\over 4}\right)
\ln{{m_b^2}\over{\mu^2_{0F}}}.
\label{yy}
\end{equation}
The resummed initial condition has a generalized exponential
structure \cite{cc},
\beq\label{deltad}
\Delta^{(D)}_N \left[ \alpha_S(\mu_{0R}^2), \mu_{0R}^2,\mu_{0F}^2, m_b^2
 \right] \, = \, 
\exp
\left\{
G^{(D)}_N \left[ \alpha_S(\mu_{0R}^2), \mu_{0R}^2,\mu_{0F}^2, m_b^2 \right]
\right\},
\eeq
where
\beq
\label{resdini}
G^{(D)}_N\left[ \alpha_S(\mu_{0R}^2), \mu_{0R}^2, \mu_{0F}^2, 
m_b^2 \right] \, = \,
\int_0^1 {dz {{z^{N-1}-1}\over{1-z}}}
\left\{\int^{\mu_{0F}^2}_{m_b^2 (1-z)^2} {{dk^2}\over {k^2}}
A\!\left[\alpha_S(k^2)\right]  
\, + \, D\!\left[\alpha_S\left(m_b^2(1-z)^2\right)\right]\right\},
\eeq
with $k^2$ and $z$ defined as in (\ref{GNC}).
As in the coefficient function, the LLs in the exponent $G_N^{(D)}$ 
are $\sim\alpha_S^n\ln^{n+1}N$,
the NLLs $\sim\alpha_S^n\ln^nN$, and so forth.
To NNLL accuracy, we need $A^{(1)}$, $A^{(2)}$ and $A^{(3)}$,
given in the previous section, and the first two coefficients of 
 \begin{equation}
D(\alpha_S) \, = \, \sum_{n=1}^{\infty}\left({{\alpha_S}\over
{\pi}}\right)^n D^{(n)},
\end{equation}
namely
\begin{eqnarray}
D^{(1)}&=&-C_F,\\
D^{(2)}&=&C_F\left[C_A\left({{55}\over{108}}-{9\over 4}\zeta(3)+
{{\zeta(2)}\over 2}\right) \, + \, {{n_f}\over {54}}\right].\label{d2}
\end{eqnarray}
Function $D(\alpha_S)$, called $H(\alpha_S)$ in \cite{cc},
 is characteristic of the fragmentation of 
heavy quarks and resums soft-gluon
radiation which is not collinear enhanced. Its ${\cal O}(\alpha_S)$ 
coefficient can be found in \cite{cc}, 
while $D^{(2)}$ can be read from the formulas in \cite{alex1}.
Moreover, as discussed in \cite{gardi1}, $D(\alpha_S)$ 
coincides with the function which resums large-angle soft radiation
in heavy-flavour decays \cite{agl,ric,ric1}. It is also equal to
function $S(\alpha_S)$, which plays the same role
in top-quark decay \cite{ccm} and massive
Deep Inelastic Scattering \cite{cm1}.

The coefficient $D^{(2)}$ depends on the renormalization
condition on the $b$ mass and Eq.~(\ref{d2}) gives its value in the on-shell
scheme. 
If $\hat m_b$ is the $b$-quark mass in another scheme, related to
the pole mass $m_b$ via
\beq
\hat{m}_b \, = 
\, \left[ 1 \, + \, \frac{\alpha_S}{\pi} \, k^{(1)} \right] \, m_b, 
\eeq
the following relation has to be
fulfilled, with the coefficients $A^{(i)}$ clearly unchanged:
\beq
\int^{\mu_{0F}^2}_{m_b^2 (1-z)^2} {{dk^2}\over {k^2}} \, A\left[ \alpha_S(k^2) \right]  
\, + \, D \left[ \alpha_S\left(m_b^2(1-z)^2\right) \right]
\, = \, 
\int^{\mu_{0F}^2}_{{\hat{m}_b}^2 (1-z)^2} {{dk^2}\over {k^2}} \, A\left[ \alpha_S(k^2) \right]  
\, + \, \hat{D} \left[ \alpha_S\left({\hat{m}_b}^2 (1-z)^2\right) \right].
\eeq
By solving the above equation order by order, one obtains the
coefficients $\hat D^{(i)}$:
\footnote{
Since $\hat{m}_b$ depends in general on a renormalization scale
$\mu_m$, one could estimate its contribution to the 
theoretical error on the cross
section by varying --- in additional to the usual factorization and 
renormalization scales --- also $\mu_m$ around $m_b$ inside a conventional
range.}
\begin{eqnarray}
\hat{D}^{(1)} &=& D^{(1)};
\\
\hat{D}^{(2)} &=& D^{(2)} \, + \, 2 \, k^{(1)} \, A^{(1)}.\label{d22}
\end{eqnarray}

As in the case of the coefficient function, 
we identify the integral $\sim dk^2A[\alpha_S(k^2)]/k^2$ 
in Eq.~(\ref{resdini}) as
a generalization of the NLO integral in Eq.~(\ref{logd}).
Function $D(\alpha_S)$ generalizes
the single logarithmic term, $\sim 1/(1-x)_+$,  
that we found in the NLO computation (\ref{heuristic2}):
\beq
- \, C_F \, \frac{\alpha_S}{\pi}  \, = \, D^{(1)} \, \frac{\alpha_S}{\pi}
\, \, \to \, \, D\left[\alpha_S\left( m_b^2(1-x)^2 \right)\right].
\eeq
Following the arguments discussed in Subsection \ref{sec41}, the
Mellin transform in Eq.~(\ref{resdini})
will be again performed exactly and the term $\sim (z^{N-1}-1)$
will not be approximated.
The subleading terms which are resummed in this way can be obtained after
expanding Eq.~(\ref{deltad}) to ${\cal O}(\alpha_S)$:
\begin{eqnarray}
\left(\Delta^{(D)}_N
\left[ \alpha_S(\mu_{0R}^2), \mu_{0R}^2, \mu_{0F}^2, m_b^2 \right]
\right)_{\alpha_S}
&=&1+{{\alpha_S(\mu_{0R}^2)}\over{\pi}} 
\int_0^1{dz} z^{N-1}\left\{ A^{(1)}
\left[{1\over{1-z}}\ln{{\mu_{0F}^2}\over{m_b^2(1-z)^2}}\right]_+\right.
\nonumber\\
&+&\left.{{D^{(1)}}\over{(1-z)_+}}\right\}\label{ddd}\\
&=& 1+{{\alpha_S (\mu_{0R}^2)C_F}\over{\pi}}  
\left[ 
\left(1-\ln{{\mu_{0F}^2}\over{m_b^2}}\right) S_1(N-1) - S_1^2(N-1)\right.
\nonumber\\
&-& S_2(N-1)\Bigg], 
\label{resdas}
\end{eqnarray}
and taking its large-$N$ limit:
\begin{eqnarray}
\left(\Delta^{(D)}_N\left[ \alpha_S(\mu_{0R}^2), \mu_{0R}^2, 
\mu_{0F}^2, m_b^2 \right]\right)_{\alpha_S}
&=&1 \ + \, {{\alpha_S(\mu_{0R}^2)\, C_F }\over{\pi}} 
\left[
- \ln^2N + \left(1 - 2\gamma_E - \ln{{\mu^2_{0F}}\over{m_b^2}}\right)\ln N
\right.
\nonumber\\
&+&\left. Y'(\mu_{0F}^2,m_b^2) \, + \, {\cal O}\left({1\over N}\right)\right],
\label{resdnlolargen}
\end{eqnarray}
with 
\begin{equation} 
Y'(\mu_{0F}^2,m_b^2) \, \equiv \,
\left(1-\ln{{\mu^2_{0F}}\over{m_b^2}}\right)\gamma_E - 
\gamma_E^2-{{\pi^2}\over 6}.
\label{y1}
\end{equation}

\subsection{Matching of resummed and NLO initial condition}

We follow a matching procedure analogous to the one for the
coefficient function in order to obtain a reliable result 
throughout the full $N$ ($x$) range.

\subsubsection{$N$-space}
We would like to write the NNLL-resummed initial condition 
of the perturbative
fragmentation function as:
\begin{eqnarray}
D^{\rm{ini,res}}_N
\left[\alpha_S(\mu_{0R}^2), \mu_{0R}^2, \mu_{0F}^2, m_b^2  \right]  
&=& K^{(D)}\left[ \alpha_S(\mu_{0R}^2), \mu_{0R}^2, \mu_{0F}^2, m_b^2 \right] 
\, \Delta^{(D)}_N
\left[\alpha_S(\mu_{0R}^2), \mu_{0R}^2, \mu_{0F}^2, m_b^2 \right] 
\nonumber\\
&+& 
d^{(D)}_N\left[\alpha_S(\mu_{0R}^2), \mu_{0R}^2, \mu_{0F}^2, 
m_b^2 \right].\label{dfin}\end{eqnarray}
The multiplying factor is obtained subtracting from the constants terms that
are present in the NLO result, the ones which have been resummed:
\begin{equation}\label{cnd}
K^{(D)}
\left[ \alpha_S(\mu_{0R}^2), \mu_{0R}^2,\mu_{0F}^2, m_b^2 \right]=
1 \, + \, {{\alpha_S(\mu_{0R}^2)\, C_F}\over{\pi}} 
Y''(\mu^2_{0F},m_b^2)
\end{equation}
where
\begin{equation} 
Y''(\mu^2_{0F},m_b^2) \, \equiv 
\, Y(\mu^2_{0F},m_b^2) \, - \, Y'(\mu^2_{0F},m_b^2)
\, = \, 1 \, + \, {3\over 4} \ln{{\mu_{0F}^2}\over{m_b^2}},
\end{equation}
and $Y(\mu^2_{0F},m_b^2)$ and $Y'(\mu^2_{0F},m_b^2)$
given in Eqs.~(\ref{yy}) and (\ref{y1}).

The remainder function for the initial condition reads:
\begin{eqnarray}\label{dnd}
d^{(D)}_N\!\left[\alpha_S(\mu_{0R}^2), \mu_{0R}^2, \mu_{0F}^2, m_b^2 \right]
\!&=& D^{\mathrm{ini}}_N
\left[\alpha_S(\mu_{0R}^2), \mu_{0R}^2, \mu_{0F}^2, m_b^2 \right]\nonumber\\
&-&
\left(K^{(D)}
\left[ \alpha_S(\mu_{0R}^2), \mu_{0R}^2,\mu_{0F}^2, m_b^2 \right]
\Delta_N^{(D)}\left[\alpha_S(\mu_{0R}^2), \mu_{0R}^2, \mu_{0F}^2, 
m_b^2 \right]\right)_{\alpha_S}\nonumber\\
&=&
\frac{ \alpha_S(\mu_{0R}^2)C_F }{\pi }
\Bigg[
- \frac{1}{2} \ln \frac{\mu_{0F}^2}{m_b^2} 
\left(\frac{1}{N+1}+\frac{1}{N}\right)
- \psi^{(0)}(N) \left(\frac{1}{N+1} + \frac{1}{N}\right)
\nonumber\\
&-& 
\gamma_E \left( \frac{1}{N} + \frac{1}{N+1} \right)
\, - \, \frac{1}{2 N}
\, + \, \frac{3}{2 (N+1)}
\, - \, \frac{1}{N^2}
\, - \, \frac{1}{(N+1)^2}
\Bigg].
\end{eqnarray}
Multiplying the NNLL-resummed coefficient function by the
DGLAP evolution operator and the 
NNLL initial condition, we obtain the moments of 
the $b$-quark 
energy distribution in $e^+e^- \, \to \, b \bar b $:
\begin{eqnarray}
\label{sigmabn}
\sigma^b_N
\!\left[\alpha_S(\mu_{0R}^2),  \alpha_S(\mu_R^2), \mu_{0R}^2, 
\mu_R^2, \mu_{0F}^2, \mu_F^2, m_b^2, m_Z^2 
\right]\!\!\!
& = &\! C_N^{\rm res} 
\!\left[ \alpha_S(\mu_R^2), \mu_R^2, \mu_F^2, m_Z^2 \right]\! 
\times\!
E_N\!\left[ \alpha_S(\mu_{0F}^2),\alpha_S(\mu_F^2) \right]\!
\!\nonumber\\
&\times&\! 
D_N^{\mathrm{ini,res}}
\left[ \alpha_S(\mu_{0R}^2), \mu_{0R}^2, \mu_{0F}^2, m_b^2 \right].
\end{eqnarray}

\subsubsection{$x$-space}
The ${\cal O}(\alpha_S)$ expansion of the resummed initial condition
in $x$-space can be read from the integrand function of Eq.~(\ref{ddd}):
\begin{equation}\label{asn}
\left
(\Delta^{(D)}\left[ x;  \alpha_S(\mu_{0R}^2), \mu_{0R}^2, \mu_{0F}^2, m_Z^2 
\right]\right)_{\alpha_S} \!=\delta(1-x) + 
\frac{\alpha_S(\mu_{0R}^2)C_F}{\pi} \, \left\{
\left[ \frac{1}{1-x}  \ln {{\mu_{0F}^2}\over{m_b^2(1-x)^2}} \right]_+
\! - \frac{1}{ \left[ 1 - x \right]_+ }\right\}.
\eeq
The constant $K^{(D)}$
is the coefficient of the $\sim\alpha_S\,\delta(1-x)$ term
in (\ref{heuristic2}), obviously equal to (\ref{cnd}). The 
remainder function can be as well obtained from Eq.~(\ref{heuristic2}) and
reads:
\begin{eqnarray}\label{dxd}
d^{(D)}\left[x; \alpha_S(\mu_{0R}^2), \mu_{0R}^2, \mu_{0F}^2, m_b^2 \right]
\!&=& D^{\mathrm{ini}}
\left[x; \alpha_S(\mu_{0R}^2), \mu_{0R}^2, \mu_{0F}^2, m_b^2 \right]\nonumber\\
&-&
\left(K^{(D)}
\!\left[\alpha_S(\mu_{0R}^2), \mu_{0R}^2, \mu_{0F}^2, m_b^2 \right]
\Delta^{(D)}\!\left[x; \alpha_S(\mu_{0R}^2), \mu_{0R}^2, \mu_{0F}^2, m_b^2
\right]\right)_{\alpha_S}\nonumber\!\\
&=&  \frac{ \alpha_S(\mu_{0R}^2)C_F }{\pi } (1+x) 
\left[
\ln(1-x) 
\, + \, \frac{1}{2}
\, - \, \frac{1}{2} \ln \frac{ \mu_{0F}^2 }{m_b^2}
\right].
\end{eqnarray}
On can check that the Mellin transform of Eq.~(\ref{dxd}) agrees with
Eq.~(\ref{dnd}).
The NNLL-resummed initial condition in $x$-space is finally given by:
\begin{eqnarray}
D^{\rm{ini,res}}
\left[x;\alpha_S(\mu_{0R}^2), \mu_{0R}^2, \mu_{0F}^2, m_b^2  \right]  
&=& K^{(D)}\left[ \alpha_S(\mu_{0R}^2), \mu_{0R}^2, \mu_{0F}^2, m_b^2 \right] 
\, \Delta^{(D)}
\left[x;\alpha_S(\mu_{0R}^2), \mu_{0R}^2, \mu_{0F}^2, m_b^2 \right] 
\nonumber\\
&+& 
d^{(D)}\left[ x; \, \alpha_S(\mu_{0R}^2), \mu_{0R}^2, \mu_{0F}^2, 
m_b^2 \right].\end{eqnarray}
\sect{Effective coupling constant}
\label{sec6}

In this section we  introduce an effective QCD
coupling constant 
$(i)$ having no Landau pole and $(ii)$ 
resumming absorptive effects related to parton branching to all orders.

\subsection{Space-like coupling constant}

If we denote by $q^{\mu}$ the typical 4-momentum entering 
in the renormalization
conditions for the QCD coupling constant
\footnote{
One can consider, for example, the symmetric point
$p_q^2 \, = \, p_{\bar{q}}^2 \, = \, p_g^2 \, = \, q^2$ 
in the $q\bar{q}g$ correlation function.}, 
the standard LO coupling constant reads: 
\begin{equation}\label{aslo}
\alpha_{S,\mathrm{LO}}(-q^2) \, = \, \frac{1}{\beta_0 
\ln \left[(-q^2-i\epsilon) /\Lambda^2\right] }
\, = \,  
\frac{1}{\beta_0 \, \left[ \, \ln (|q^2| /\Lambda^2)
\, - \, i \pi \, \Theta(q^2) \, \right]}.
\end{equation}
There is a minus sign in front of the momentum squared $q^2$, because of
the opening of decay channels for the gluon ($g\to q\bar q$, $g\to gg$)
in the time-like region 
$q^2 \, > \, 0$.
In order to have a renormalized {\it real} $\alpha_S$, one
usually considers a space-like configuration of the reference momenta:
$q^2 \, < \, 0$.
To avoid explicit minus signs, the LO expression of $\alpha_S$
is usually written as:
\begin{equation}
\label{start}
\alpha_{S,\mathrm{LO}}(Q^2) \, = \, \frac{1}{\beta_0 \ln (Q^2/\Lambda^2)},
\end{equation}
where $Q^2 \, \equiv \, - \, q^2 \, > \, 0$ in the
space-like region.
The specific properties of the QCD coupling constant involved in soft-gluon
resummation in $e^+e^-$ annihilation are related to:
\begin{enumerate}
\item
an integration up to small momentum scales of the coupling constant, 
because of multiple parton radiation, giving rise to 
sub-jets with arbitrarily small masses. In the resummed expressions,
(\ref{GNC}) and (\ref{resdini}),
one can indeed observe that the scale of $\alpha_S$ approaches zero
once $x\to 1$.
\item
the kinematical configurations
are always time-like, as we are considering multiple emissions in the
final-state of $e^+e^-$ processes.
\end{enumerate}
Let us deal with the above issues by discussing
first the analyticity properties of the standard coupling constant
(\ref{start}).
$\alpha_S(Q^2)$  exhibits a cut for $Q^2 \, < \, 0$, 
associated with the branching of a time-like 
gluon ($q^2 \, > \, 0$) into `physical' states, 
and the Landau pole for $Q^2 \, = \, \Lambda^2$.
While the former singularity has a clear meaning, 
the Landau pole is not physical and it just reflects the unreliability
of the perturbative expansion for $Q^2\approx \Lambda^2$.

It was therefore suggested than one can replace 
the usual expression (\ref{start}) with an analytic coupling 
$\bar\alpha_S(Q^2)$, which 
has the same discontinuity as $\alpha_S(Q^2)$
along the cut $Q^2 \, \le \, 0$, but is analytic elsewhere
in the complex plane. As in \cite{ermolaev,shirkov,stefanis},
we write the analytic coupling constant
$\bar\alpha_S(Q^2)$ using the following dispersion relation:
\begin{equation}
\bar\alpha_S(Q^2)= 
\frac{1}{2\pi i}
\int_0^{\infty}  \, \frac{ds}{s+Q^2} \, 
{\rm Disc}_s \, \alpha_S(-s), 
\label{space}
\end{equation}
where the discontinuity is defined as:
\begin{equation}
{\rm Disc}_s F(s) =  \lim_{\epsilon \, \to \, 0^+} 
\left[F(s+i\epsilon)- F(s-i\epsilon)\right].
\end{equation}
Eq.~(\ref{space}) holds for $Q^2 > 0$, i.e. in the space-like region
$q^2<0$: as in \cite{shirkov}, we shall refer to it as our
`space-like' analytic coupling constant. 
For $Q^2<0$, the integrand function in Eq.~(\ref{space}) presents a pole
in the domain of integration. However, we can still give sense to 
Eq.~(\ref{space}) for negative values of $Q^2$, introducing a small imaginary 
part: $Q^2\to Q^2+i\epsilon$.

Inserting in (\ref{space}) the LO expression (\ref{start}), we get the LO
space-like coupling constant:
\begin{equation}
\bar\alpha_S(Q^2) \, = \,{1\over{\beta_0}}\left[{1\over{\ln(Q^2/\Lambda^2)}}-
{{\Lambda^2}\over{Q^2-\Lambda^2}}\right].
\label{spacelo}
\end{equation}
In (\ref{spacelo}) the Landau pole has been 
subtracted by a power-suppressed term, relevant at small $Q^2$
and negligible at large $Q^2 \, \gg \,\Lambda^2$, 
where $\bar\alpha_S(Q^2)$ still
exhibits the same behaviour as $\alpha_S(Q^2)$. 
Likewise, including in the integrand function of (\ref{space}) the higher-order
expressions of $\alpha_S(-s)$, one can get the space-like
$\bar\alpha_S(Q^2)$ to higher accuracy.

\subsection{Time-like coupling constant including absorptive effects}

Turning back to our calculation, we have resummed soft and/or
collinear multiple radiation in the final state of $e^+e^-$ annihilation, 
i.e. a time-like parton cascade. 
Also, in our resummed expressions, (\ref{GNC}) and (\ref{resdini}),
the coupling 
constant is evaluated at a scale $k^2$, which is roughly
the transverse momentum of the emitted parton
with respect to the radiating one.
In Ref.~\cite{amati}, it was in fact shown that, in the framework of
resummed calculations, the momentum-independent coupling constant is to
be replaced by the following integral over the discontinuity of the gluon 
propagator: 
\begin{equation}
\alpha_S\to \frac{i}{2 \pi} \, \int_0^{k^2} d s 
\ {\rm Disc}_s\  \frac{\alpha_S(-s) }{ s }.
\label{ask}
\end{equation}
The integral (\ref{ask}) is typically performed neglecting the imaginary part,
$\sim i\pi$, in the denominator of $\alpha_S(-s)$ (see 
for example Eq.~(\ref{aslo})),
i.e. assuming
\begin{equation}\label{ppi}
\ln{{|s|}\over{\Lambda^2}}\gg \pi
\end{equation}
in the integrand function of (\ref{ask}).
As a result, the integral (\ref{ask}) turns out to be  
approximately equal to $\alpha_S$ evaluated at the upper integration limit:
\begin{equation}
\frac{i}{2 \pi} \, \int_0^{k^2} d s 
\ {\rm Disc}_s\  \frac{\alpha_S(-s) }{ s }\, \simeq\, \alpha_S(k^2).
\label{assk}
\end{equation}
In our analysis, we wish 
to go beyond the assumption (\ref{ppi}) and account for the
terms $\sim i\pi$ in the denominator of $\alpha_S$; 
this way, as pointed out in \cite{agl}, one includes
absorptive effects due to gluon branching, that are important especially
in the infrared region.
The new feature of our model is the fact that
we avoid the Landau pole in the integral by using in the integrand
function the space-like analytic coupling constant $\bar\alpha_S(-s)$,
just defined in Eq.~(\ref{space}).

As a result, our 
model consists in using the following `time-like' effective
coupling constant: 
\begin{equation}
\tilde\alpha_S(k^2) = \frac{i}{2 \pi} \, \int_0^{k^2} d s 
\ {\rm Disc}_s\  \frac{ \bar\alpha_S(-s) }{ s }.
\label{time}
\end{equation}
In fact, Eq.~(\ref{time}) makes sense only for $k^2>0$: the above integral 
would be zero for negative values of $k^2$.
We also remark a difference in our notation:
the effective coupling $\tilde{\alpha}_S(k^2)$ is 
function of the square of a four-momentum $k^2$; the standard 
$\alpha_S(-q^2)$ in Eq.~(\ref{start}) is instead function of minus a
squared  four-momentum.
At large $k^2$, $\tilde\alpha_S(k^2)$ will be roughly equivalent to the
standard $\alpha_S(k^2)$; at small $k^2$ it will include non-perturbative
power-suppressed
effects. The goal of this paper is precisely to investigate
whether including non-perturbative corrections using
Eq.~(\ref{time}) everywhere in our calculation is suitable 
to reproduce the experimental data on $B$-hadron production,
without adding any further hadronization model.

Inserting Eqs.~(\ref{spacelo}) in the integrand function
of (\ref{time}), we obtain the LO time-like analytic coupling constant:
\begin{equation}
\tilde\alpha_{S,\mathrm{LO}}(k^2) \, = \, 
\frac{1}{2\pi i \beta_0}
\left[\ln\left( \ln \frac{k^2}{\Lambda^2} + i \pi \right)
-\ln\left( \ln \frac{k^2}{\Lambda^2} - i \pi \right)\right].
\label{timelo}
\end{equation}
Likewise, starting from the NLO $\alpha_S(Q^2)$
in Eq.~(\ref{alphanlo}), we obtain the NLO time-like coupling constant: 
\begin{eqnarray}
\tilde\alpha_{S,\mathrm{NLO}}(k^2) &=&\tilde\alpha_{S,\mathrm{LO}}(k^2)+
\frac{\beta_1}{\beta_0^3}
\frac{1}{2\pi i}
\left[\frac{ \ln\left[ \ln (k^2/\Lambda^2) + i \pi \right] 
+1}{\ln(k^2/\Lambda^2)+i\pi}-\frac{\ln\left[\ln(k^2/\Lambda^2) 
-i\pi\right]+1}{\ln (k^2/\Lambda^2) - i \pi } \right].
\label{timenlo}
\end{eqnarray}
At NNLO, the standard coupling constant reads:
\begin{eqnarray}
\alpha_S(k^2)&=&{1\over{\beta_0\ln(k^2/\Lambda^2)}}
\left\{ 1-{{\beta_1}\over{\beta_0^2}}
{{\ln[\ln(k^2/\Lambda^2)]}\over{\ln(k^2/\Lambda^2)}}
+{{\beta_1^2}\over{\beta_0^4}}
{{\ln^2[\ln(k^2/\Lambda^2)]-\ln[\ln(k^2/\Lambda^2)]-1}
\over{\ln^2(k^2/\Lambda^2)}}\right.\nonumber\\
&+&\left.{{\beta_2}\over{\beta_0^3}}
{1\over{\ln^2(k^2/\Lambda^2)}}\right\},
\end{eqnarray}
and its time-like analytic counterpart:
\begin{eqnarray}
\tilde\alpha_{S,\mathrm{NNLO}}(k^2) &=& \tilde\alpha_{S,\mathrm{NLO}}(k^2)
-\frac{\beta_1^2}{4\pi i \beta_0^5}
\left\{\frac{\ln^2\left[ \ln (k^2/\Lambda^2) + i \pi \right]}
{\left[ \ln (k^2/\Lambda^2) + i \pi \right]^2}
-\frac{ \ln^2\left[ \ln (k^2/\Lambda^2) - i \pi \right] }
             { \left[ \ln (k^2/\Lambda^2) - i \pi \right]^2 }
\right\}
\nonumber\\
& +& \frac{\beta_1^2- \beta_0 \beta_2 }{4\pi i \beta_0^5}
\left\{\frac{ 1 }{ \left[\ln (k^2/\Lambda^2) + i \pi \right]^2 }
 - \frac{ 1 }{ \left[ \ln (k^2/\Lambda^2) - i \pi \right]^2 }
\right\}.
\label{timennlo}
\end{eqnarray}
In (\ref{timennlo}) we have also included the third coefficient of
the $\beta$-function:
\begin{equation}
\beta_2 \, = \, {1\over{64\pi^3}}\left[{{2857}\over{54}}C_A^3-
\left({{1415}\over{54}}C_A^2+{{205}\over{18}}C_AC_F-C_F^2\right)n_f
+\left({{79}\over{54}}C_A+{{11}\over 9}C_F\right)n_f^2\right]. 
\end{equation}
In Eqs.~(\ref{timelo}), (\ref{timenlo}) and (\ref{timennlo}) we have
used a complex notation, which yields quite compact expressions for
the time-like coupling constant. However, we can rearrange the above
equations and express
$\tilde\alpha_S(k^2)$ as a real function of $k^2$,  as done in \cite{ric}.

Before closing this section, we point out a few more issues. 
Expanding $\tilde\alpha_S(k^2)$ for $\ln(k^2/\Lambda^2)\gg \pi$,
it is possible to relate standard and analytic time-like coupling constants:
\begin{equation}
\tilde\alpha_S(k^2) \, = \, \alpha_S(k^2) 
\, - \, \frac{\left(\pi\beta_0\right)^2}{3} \, \alpha_S^3(k^2)
\, + \, {\cal O}(\alpha_S^4).
\label{atas}
\end{equation}
We also need 
to modify the matching condition for the strong coupling constant 
when running from $n_f$ to $n_{f-1}$ active flavours.
In terms of the standard $\alpha_S(k^2)$, it reads \cite{bern}:
\begin{equation}
\alpha_{S(n_f)}(\bar{m}^2_q) \, = \, \alpha_{S(n_f-1)}(\bar{m}^2_q) - 
{{11}\over{72\pi^2}}
\alpha^3_{S(n_f-1)}(\bar{m}^2_q) + {\cal O}(\alpha_S^4),
\label{nf}
\end{equation}
where $\bar{m}_q$ is the running quark mass in the \msbar renormalization
scheme, i.e. $\bar{m}_q \, = \, \bar{m}_q^{\overline{\mathrm{MS}}}(\bar{m}_q)$.
When using the analytic effective coupling $\tilde\alpha_S$, 
we shall have to modify Eq.~(\ref{nf}) according to Eq.~(\ref{atas}),
observing that $\beta_0$, given in Eq.~(\ref{b0b1}) depends on $n_f$.
We obtain:
\begin{equation}
\tilde\alpha_{S(n_f)}(\bar{m}_q^2) \, 
= \, \tilde\alpha_{S(n_f-1)}(\bar{m}_q^2)-
\left({{11}\over{72\pi^2}} - {{17-n_f}\over{54}}\right)
\tilde\alpha^3_{S(n_f-1)}
(\bar{m}_q^2)+{\cal O}(\alpha^4_S).
\label{nnf}
\end{equation}
Of course,
Eqs.~(\ref{nf}) and (\ref{nnf}) can be also expressed in terms of the
pole quark masses.

\sect{Modelling non-perturbative corrections}
\label{sec7}

In this section we describe our  model for non-perturbative
effects in bottom-quark fragmentation, based on the effective QCD
coupling constant considered above.
Some properties of the model have already been anticipated in the
previous section; here we present a systematic discussion.

We shall use the effective coupling constant 
$\tilde{\alpha}_S(k^2)$ defined through Eq.~(\ref{time}) 
in place of the standard one, in order to
include power corrections.
The dominant non-perturbative effects in $b$-fragmentation
occur for 
\beq
1-\frac{\mu}{m_b}\lsim x\lsim 1,
\eeq
where $\mu$ is of the order of the QCD scale, i.e. 
$\mu\sim {\cal O}(\Lambda)$.
Within our model, such contributions are
associated with soft interactions of the $b$ quark in the
fragmentation into a $B$ hadron
and are analogous to the well-known Fermi motion
of a decaying $b$ quark inside a $B$.
In the perturbative fragmentation approach, such effects become relevant
when we evaluate at large $x$ 
the quantities 
\beq
C=m_Z\sqrt{1-x}\ \ \ {\mathrm and}\ \ \ 
S=m_b(1-x),
\eeq
whose squares are the limits of 
the $dk^2/k^2$ integration in the resummed coefficient
function (\ref{GNC})
and initial condition (\ref{resdini}), respectively,
as well as the scales of functions $B(\alpha_S)$ and $D(\alpha_S)$.
It is interesting to evaluate $C$, $S$ and the corresponding values
of $\tilde\alpha_S$ for $x=0.8$, since, as we shall show in the following
section, the $B$-hadron spectrum in 
$e^+e^-$ annihilation is peaked about this value of $x$.
We find:\footnote{For $x=0.9$, the corresponding numbers are:
$C\simeq 30$~GeV, $\tilde\alpha_S(C^2)\simeq 0.14$, $S\simeq 0.5$~GeV,
$\tilde\alpha_S(S^2)\simeq 0.44$.} 
\begin{equation}\label{scales}
C\simeq 40~\mathrm{GeV}\ \ \ ,\ \ \  \tilde\alpha_S(C^2)\simeq 0.13
\ \ \ \ ;\ \ \ 
S\simeq 1~\mathrm{GeV}\ \ \ ,\ \ \  \tilde\alpha_S(S^2)\simeq 0.33.
\end{equation}
It is therefore clear that non-perturbative effects are more relevant
in the initial condition (\ref{resdini}), where $S$ plays a role, rather 
than in the coefficient function (\ref{GNC}), depending on $C$. 
In order to deal with such effects, 
we shall insert   $\tilde{\alpha}_S$ in place of the standard
$\alpha_S$ in the resummed initial condition (\ref{resdini})
and, as part of our model, we shall 
perform the Mellin and inverse Mellin transforms exactly. 
As discussed in Subsection \ref{sec41}, the issue of the integration 
over $z$ when 
using the effective coupling is delicate and does deserve a deeper 
investigation in the next future. 
However, it was pointed out in \cite{Aglietti:2004fz} that, in order to
include the power-suppressed corrections originated by the effective
coupling, $\sim {\cal O} (1/N)$ or $\sim{\cal O}(1-x)$, 
performing the $z$-integration exactly is necessary.  
In fact, an approximate Mellin transform, using the step function (\ref{step}) 
or the formulas beyond NLL in \cite{defl},
would suppress most of such effects \cite{Aglietti:2004fz}.
The Mellin transforms were performed exactly even in 
Ref.~\cite{noi4}, where the effective coupling was used to describe 
non-perturbative effects in $B$-meson decays.

In the coefficient function, the use 
of $\tilde\alpha_S(k^2)$ and the exactness of the Mellin transform
are less crucial than in the initial condition of the
perturbative fragmentation function.
In any case, for practical convenience,
we shall still use the effective coupling constant and perform the
$z$-integration exactly.  Besides, $\tilde\alpha_S(k^2)$ will be
employed everywhere in our calculation, including the constant 
terms and the remainder functions presented in Eqs.~(\ref{dnc}), (\ref{dxc}),
(\ref{dnd}) and (\ref{dxd}).

We also remark that that there is not a unique way to construct a model 
based on the analytic coupling constant.
In fact, we have to make two choices:
\begin{enumerate}
\item
we can include the absorptive effects, which are always present in 
time-like kinematics, and use $\tilde\alpha_S(k^2)$; alternatively,
we do not include such effects and employ the space-like
$\bar{\alpha}_S(k^2)$;
\item
we can perform a 
power expansion of the higher orders or not.

The latest point deserves some more comments.
By `power expansion' we mean that higher orders have in front
a power of the effective coupling constant:
\beq
\label{power_exp}
\tilde{\alpha}_S^n(k^2)=
\left[\frac{i}{2 \pi} \, \int_0^{k^2} d s 
\ {\rm Disc}_s\  \frac{ \bar\alpha_S(-s) }{ s }\right]^n.
\eeq
The $n$-th power is taken after the discontinuity of $\bar\alpha_S(-s)$
is computed and the integral over the gluon virtuality $s$
is performed.
Adopting the `non-power expansion' choice, as
originally proposed in \cite{shirkov}, 
consists instead
in evaluating first the discontinuity of $\bar\alpha_S^n(-s)$,
and then integrating over $s$. 
Formally, $\tilde{\alpha}^n_S(k^2)$ will have to be replaced according to:
\begin{equation}
\label{timen}
\tilde{\alpha}_S^n(k^2) ~~ \to \\
\widetilde{\alpha_S^n}(k^2) \, \equiv \, \frac{i}{2 \pi} \, \int_0^{k^2} d s 
\, {\rm Disc}_s \,  \frac{ \bar\alpha_S^n(-s) }{ s }.
\end{equation}
Unlike (\ref{power_exp}), Eq.~(\ref{timen}) is linear in the discontinuity.
\end{enumerate}
We have followed an empirical criterion, leaving to the future
the task of a theoretical justification:
we select the model which gives spectra closer to the data,
without adding any further non-perturbative fragmentation
function.
We can anticipate that 
have found that the power expansion of the coupling constant 
which includes
the absorptive effects, i.e. $\tilde{\alpha}_S(k^2)$, 
leads to the best description 
of the experimental data.
To be more general, we stress that our choices in modelling non-perturbative
effects are partly related to the accuracy of the perturbative
resummed calculation that we are using.
Considering, for example, function $A[\alpha_S(k^2)]$, we have used its
expansion to the third order and the NNLO $\tilde\alpha_S(k^2)$.
However, if we were to know function
$A$ to a different level of approximation or, in principle, even to any order, 
a different effective coupling constant may still yield
the same $A[\alpha_S(k^2)]$ and the same $B$-hadron spectrum.

Since Eq.~(\ref{atas}) is basically a change of scheme
for the coupling constant, the coefficients of the terms of
${\cal O} (\alpha_S^3)$ in functions $A(\alpha_S)$, $B(\alpha_S)$ and
$D(\alpha_S)$, appearing in the
resummed expressions (\ref{GNC}) and (\ref{resdini}), are
to be modified. In our NNLL approximation, we 
need to replace $A^{(3)}$ according to: 
\begin{equation}
A^{(3)} \, \to \, \tilde A^{(3)} \, = \, A^{(3)} \, + 
\, {{(\pi\beta_0)^2}\over 3}A^{(1)},
\label{a3}
\end{equation}
where we have made use of Eq.~(\ref{atas}).
The other coefficients entering in functions $A(\alpha_S)$,
$B(\alpha_S)$ and $D(\alpha_S)$ at NNLL are instead left unchanged when
replacing $\alpha_S(k^2)$ with $\tilde\alpha_S(k^2)$.

Let us now discuss another delicate point of our model.
In standard resummations,
which use the standard coupling constant and resum only the
logarithms of the Mellin variable $N$, there
is a clear counting of the terms in $\alpha_S(k^2)$ 
which are to be kept or dropped.
For example, in the NLL approximation:
\beq
A\left[\alpha_S(k^2)\right] 
\, \to \, 
        A^{(1)} \, \frac{\alpha_{S,\mathrm{NLO}}(k^2)}{\pi} 
\, + \, A^{(2)} \, \frac{\alpha^2_{S,\mathrm{LO}}(k^2)}{\pi^2}. 
\eeq
That is because
\beq
\alpha^2_{S,\mathrm{LO}} (k^2) \sim \frac{1}{\ln^2 (k^2/\Lambda^2)},
\eeq
and therefore it has, for $k^2\to\infty$, 
the same asymptotic behavior
as $\alpha_{S,\mathrm{NLO}}(k^2)$, which contains a correction term
behaving like $\ln(\ln (k^2/\Lambda^2))/\ln^2 (k^2/\Lambda^2)\sim 
1/\ln^2 (k^2/\Lambda^2)$.
In the standard resummation scheme --- which is a kind of minimal scheme ---
only logarithms are resummed and the asymptotic expansion of the coupling
constant is used to organize the series of the infrared logarithms.
On the other hand, the large-$k^2$ expansion of our time-like 
$\tilde\alpha_S(k^2)$ is 
much more complicated; even the lowest order
$\tilde{\alpha}_{S,\mathrm{LO}}(k^2)$, proportional to $1/\beta_0$, presents
terms of any order for $k^2 \, \to \, \infty$:
\beq
\tilde{\alpha}_{S,\mathrm{LO}}(k^2) \sim
\frac{c_1}{\ln (k^2/\Lambda^2)} \, + \, 
\frac{c_3}{\ln^3 (k^2/\Lambda^2)} \, + \, \frac{c_5}{\ln^5 (k^2/\Lambda^2)}
\, + \, \cdots.
\eeq
In fact, we observe that: ($i$) in our model we would like 
to include as many contributions as possible;
($ii$) there is no real reason to neglect higher-order corrections to 
$\tilde\alpha_S(k^2)$, proportional to
$\beta_1$, $\beta_2$ and so on. Therefore, we find it safe using
$\tilde{\alpha}_{S,\mathrm{NNLO}}(k^2)$
everywhere in the resummed expressions. For example:
\beq\label{annlo}
\tilde{A}[\tilde{\alpha_S}(k^2)] ~ \to ~
        {{A^{(1)}}\over\pi} \, \tilde{\alpha}_{S,\mathrm{NNLO}}(k^2)
\, + \, {{A^{(2)}}\over{\pi^2}} \, \tilde{\alpha}^2_{S,\mathrm{NNLO}}(k^2)
\, + \, {{\tilde{A}^{(3)}}\over{\pi^3}} \, \tilde{\alpha}^3_{S,\mathrm{NNLO}}
(k^2).
\eeq
In this way we include all the logarithmic terms
of the standard NNLL resummation, plus some subleading contributions.
Another possibility might have been:
\beq\label{a33}
\tilde{A}(\tilde{\alpha_S}(k^2)) ~ \to ~
        {{A^{(1)}}\over{\pi}} \, \tilde{\alpha}_{S,\mathrm{NNLO}}(k^2)
\, + \, {{A^{(2)}}\over{\pi^2}} \, \tilde{\alpha}^2_{S,\mathrm{NLO}}(k^2)
\, + \, {{\tilde{A}^{(3)}}\over{\pi^3}} \, \tilde{\alpha}^3_{S,\mathrm{LO}}(k^2).
\eeq
However, at small $k^2$, 
where we are mostly sensitive to non-perturbative effects,
higher-order corrections to the 
time-like coupling constant are negative and sizable.
Comparing, in fact, Eqs.~(\ref{timelo}), (\ref{timenlo}) and (\ref{timennlo})
for low values of $k^2$, we find:
\begin{equation}
\tilde\alpha_{S,\rm NNLO}(k^2)<\tilde\alpha_{S,\rm NLO}(k^2)
<\tilde\alpha_{S,\rm LO}(k^2).\end{equation}
Hence, if we used Eq.~(\ref{a33}) rather than Eq.~(\ref{annlo}),
the term proportional to $\tilde{A}^{(3)}$ would be enhanced
at large $x$. We can anticipate that 
this would worsen the comparison with the experimental data.

\sect{Phenomenology --- $x$-space}
\label{sec8}

In this section 
we present results on the $B$-hadron energy spectrum using the 
resummed partonic
calculation based on the perturbative fragmentation formalism,
and modelling non-perturbative effects by means of the time-like
effective coupling constant.
The basic assumption of our model is that,
whenever we use $\tilde{\alpha}_S(k^2)$ instead of $\alpha_S(k^2)$, 
the $b$-quark energy fraction $x$ will have to be replaced by its hadron-level 
counterpart $x_B$:
\begin{equation}
x_B \, \equiv \, {{2p_B\cdot q}\over{m_Z^2}},
\end{equation}
with $p_B$ being the momentum of a $b$-flavoured hadron produced
in $e^+e^-$ annihilation.

The $B$-hadron spectrum in moment space 
can be obtained from Eq.~(\ref{sigmabn}), after replacing the
standard coupling constant with the analytic time-like one:
\begin{eqnarray}
\label{sigmabhad}
\sigma^{(B)}_N( \mu_R^2, \mu_{0R}^2, \mu_{0F}^2, \mu_F^2, m_b^2, m_Z^2 )
&=& C_N^{\rm res}
\left[ \tilde\alpha_S(\mu_R^2), \mu_R^2, \mu_F^2, m_Z^2 \right] \times\
E_N\left[ \tilde{\alpha}_S(\mu_{0F}^2), \tilde{\alpha}_S(\mu_F^2) \right]
\nonumber\\
&\times &
D_N^{\mathrm{ini,res}} 
\left[ \tilde{\alpha}_S(\mu_{0R}^2), \mu_{0R}^2, \mu_{0F}^2,m_b^2 \right].
\end{eqnarray}
In Eq.~(\ref{sigmabhad}) we assume that the coefficient
function and the initial condition are resummed in the NNLL approximation,
and that the NNLO analytic coupling constant (\ref{timennlo}) is used.
In order to recover the $x$-space results, we perform
the inverse Mellin transform of Eq.~(\ref{sigmabhad}):
\beq\label{inverse}
\sigma^{(B)}
\left( x_B; \,  \mu_R^2, \mu_{0R}^2, \mu_{0F}^2, \mu_F^2, m_b^2, m_Z^2 \right)
\, = \, \int_{c-i\infty}^{c+i\infty}
\frac{dN}{2\pi i} x_B^{-N} 
\, \sigma^{(B)}_N( \mu_R^2, \mu_{0R}^2, \mu_{0F}^2, \mu_F^2, m_b^2, m_Z^2 ),
\eeq
where $c$ is a positive constant.
We point out 
that the inverse transform (\ref{inverse})
is computed in the standard mathematical way:
no prescription to deal with the Landau pole, such as the 
minimal prescription \cite{min}, is needed. 
That is because our effective coupling $\tilde\alpha_S(k^2)$
does not present the Landau pole any longer. 
Since the inverse transform is made in a numerical way,
we have checked that the results are stable with respect to change 
of the integration contour, i.e. of $c$.

\subsection{$B$-hadron spectrum and comparison with experimental data}

Following \cite{drol}, we consider data from SLD \cite{sld}, ALEPH 
\cite{aleph} and OPAL \cite{opal} collaborations at the $Z^0$ pole. 
ALEPH reconstructed only $B$ mesons, while the SLD and OPAL samples also 
contain a small fraction of $b$-flavoured baryons 
\footnote{
A naive estimation based on the $1/N_c$ expansion, where $N_c$ is the 
colour number, would predict a fraction of about
$1/N_c^2\approx \,10 \%$ $b$-flavored hyperions compared to $B$ mesons.
}. In principle,
one should consider such data separately, as the hadron content is different;
also, 
from the theoretical viewpoint, as pointed out in the introduction, there is no
real reason why the same model should describe both meson and baryon 
production.
However, the SLD and OPAL samples are inclusive and 
the baryons are anyway very little.  
In fact, Ref.~\cite{drol} found that it is possible to describe all data
points fitting the Kartvelishvili non-perturbative fragmentation model
\cite{kart} or the cluster and string models implemented in
HERWIG \cite{herwig} and PYTHIA \cite{pythia}.
It is therefore reasonable using the same hadronization model, in our
case the effective coupling constant, for the comparison with
all three experiments, and investigating how it fares against 
all data and against each experimental sample. As in the previous
analyses, when doing the comparison, we neglect the
correlations between the data points and sum the experimental systematic and 
statistical errors in quadrature.

Unlike Refs.~\cite{cm,ccm,corc,drol}, we shall
not perform a fit to the data, since we do not have any tunable parameter
in our model, apart of course from the ones contained in our perturbative
calculation.
Rather, we shall investigate the theoretical uncertainty on our
prediction, by varying the parameters in our computation, such
as scales and quark masses.
The default values of our parameters will be: 
\beq
\mu_R \, = \, \mu_F \, = \, m_Z; ~~~~~\mu_{0R} \, = \, \mu_{0F} \,= \, m_b,
\eeq 
where $\mu_R$ and $\mu_F$ are the renormalization and factorization scales
in the coefficient function (\ref{cfin}), respectively, and $\mu_{0R}$ and
$\mu_{0F}$ in the initial condition (\ref{dfin}).
Consistently with the Particle Data Group \cite{pdg}, we set
$m_Z \, = \, 91.19$~GeV and $\alpha_S(m_Z^2) \, = \, 0.119$ 
\footnote{This choice corresponds to $\tilde\alpha_S(m_Z^2) \, = \, 0.117$, 
as can be obtained from Eq.~(\ref{atas}).}. 
For the \msbar quark masses, entering in the matching condition (\ref{nf}), we
choose:
$\bar{m}_b \, = \, 4.2$~GeV,
$\bar{m}_c \, = \, 1.25$~GeV, 
$\bar{m}_s \, = \, 0.1$~GeV, 
while the up- and down-quark masses will be neglected. 
The choice of $m_b$, the $b$-quark mass entering in the
initial condition of the perturbative fragmentation function,
deserves some extra comment.
In fact, using the pole or the \msbar $b$ mass in the initial condition 
is irrelevant at NLO and when resumming threshold logarithms up to 
NLL accuracy.
Beyond such approximations, the choice of the renormalization scheme makes
a difference. Refs.~\cite{alex1,alex2}, which calculate the initial condition
to NNLO,  use the pole mass in their
computation. Although in the matching we are using the NLO initial condition, 
we are still relying on Ref.~\cite{alex1} for
the coefficient $D^{(2)}$ in the NNLL resummation (\ref{resdini}),
and therefore
we should use the pole mass as well. However, since we are aiming at predicting
the $B$-hadron energy distribution, it is not uniquely determined whether,
after using the analytic coupling constant to include power-suppressed
effects, $m_b$ should be the $b$-quark or the $B$-hadron mass. 
We believe that it is safe adopting a quite conservative choice, i.e. 
$4.7~\mathrm{GeV} \, < \, m_b \, < \, 5.3~\mathrm{GeV}$, 
which includes the present
estimations for the $b$ pole mass as well as $b$-flavoured hadron masses
\cite{pdg}. Our default value will be $m_b \, = \, 5$~GeV.

Before presenting our results, we point out that 
this 
kind of systematic analysis of the theoretical uncertainty was not performed,
e.g., in Refs.~\cite{ccm,corc,drol}.
In fact, when using a hadronization model, 
the fitting procedure would possibly adjust the free parameters to
reproduce the data, even when varying the inputs of the
parton-level computation. As long as a reliable hadronization model is used,
changing the perturbative quantities will just lead to different
best-fit parameters in the non-perturbative fragmentation function.
\begin{figure}[t]
\centerline{\resizebox{0.65\textwidth}{!}{\includegraphics{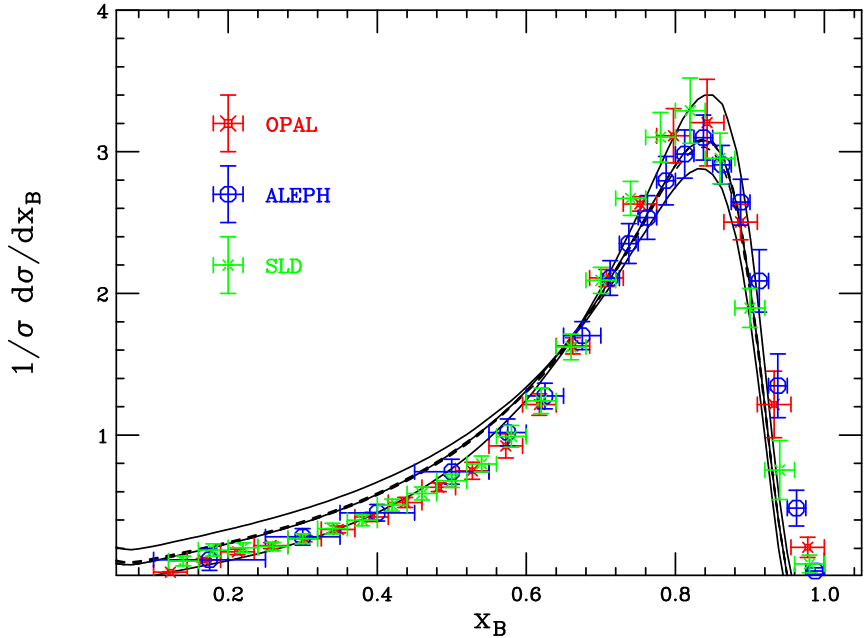}}}
\caption{\small $B$-hadron spectrum yielded by our resummed 
calculation using the analytic time-like 
coupling constant $\tilde\alpha_S(k^2)$
to model non-perturbative effects.
We investigate the dependence on the factorization scales.
Solid lines: $\mu_{0F} \, = \, m_b/2$, $m_b$ and $2m_b$;
dashed lines: $\mu_F \, = \, m_Z/2$, $m_Z$ and $2m_Z$. 
The other parameters are kept to their default values.}
\label{plotmu}
\end{figure}
\begin{figure}
\centerline{\resizebox{0.65\textwidth}{!}{\includegraphics{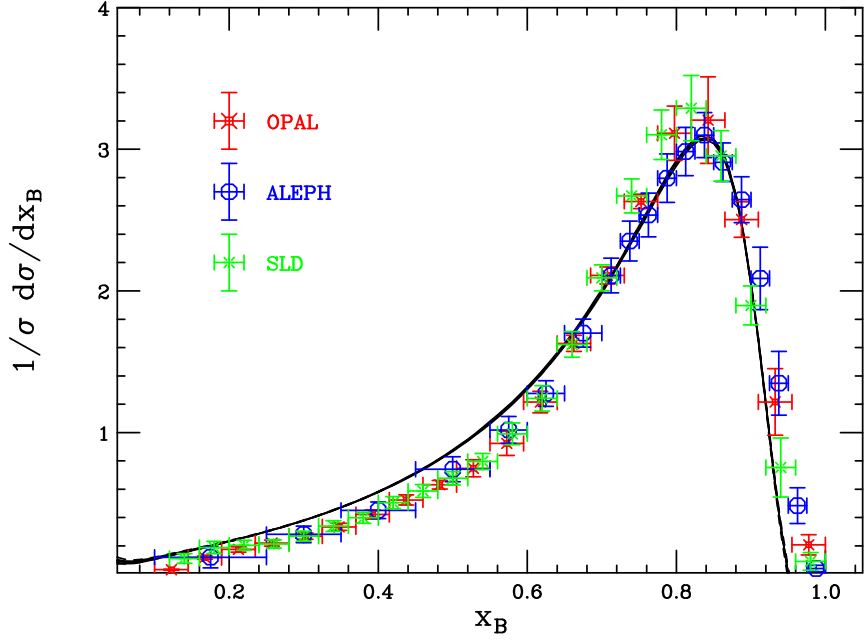}}}
\caption{\small 
Dependence on the renormalization scales $\mu_R$ and $\mu_{0R}$. 
Solid lines: $\mu_R$ varied between $m_Z/2$ and $2 \, m_Z$;
dashed lines:  $\mu_{0R}$ between $m_b/2$ and $2 \, m_b$.}
\label{plotmur}
\end{figure}
\begin{figure}
\centerline{\resizebox{0.65\textwidth}{!}{\includegraphics{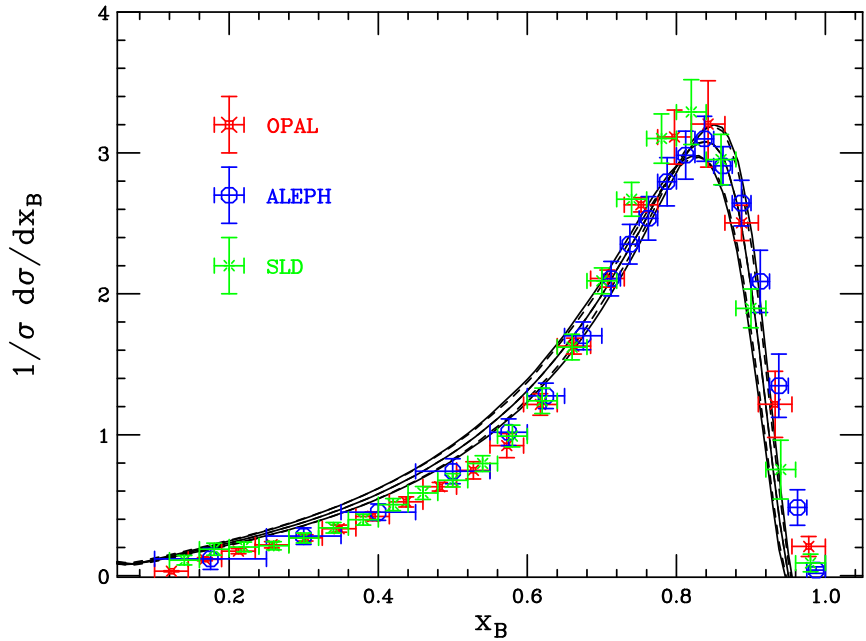}}}
\caption{\small Dependence of the $B$-energy distribution on
$\alpha_S(m_Z^2)$ and on $m_b$.
Solid lines: 
standard coupling $\alpha_S(m_Z^2)$ varied from $0.117$ to $0.121$;
Dashed lines: $b$ pole mass varied from $4.7$ to $5.3$ GeV.}
\label{plotalpha}
.\end{figure}
\par 
In Fig.~\ref{plotmu} we show the data points and our prediction,
investigating its dependence on the factorization scales
$\mu_F$ and $\mu_{0F}$. 
In Fig.~\ref{plotmur} we look instead at the dependence
on the renormalization scales $\mu_R$ and $\mu_{0R}$.
For $\mu_R$ and $\mu_F$ we choose the values $m_Z/2$, $m_Z$ and $2m_Z$;
for $\mu_{0R}$ and $\mu_{0F}$, $m_b/2$, $m_b$ and $2m_b$. 
We vary each scale separately, 
keeping all other quantities to their default values,
in order to avoid an excessive number of runs.
In Fig.~\ref{plotmu} we notice that the dependence of our prediction on
$\mu_{0F}$, the factorization scale entering in the
initial condition of the perturbative fragmentation function,
is rather large, especially at small $x_B$ and around the peak. 
At small $x_B$, the spectrum varies up to a factor of 2 if 
$\mu_{0F}$ changes from $m_b/2$ to $2m_b$; around the peak the impact
of the choice of $\mu_{0F}$ is about 20\%.
The effect of the value chosen for $\mu_F$ is instead pretty small, and well
within the band associated with the variation of $\mu_{0F}$.
A possible explanation of the fairly large dependence on $\mu_{0F}$
could be the fact that, although we are working in the NNLL approximation, 
we are still matching the resummation to the NLO exact result,
and not to the NNLO one. This generates a mismatch between the NNLL
terms in the resummed expressions ($\sim \alpha_S^2\ln N$,
$\alpha_S^3\ln^2N,\dots$) and the remainder functions, 
given in Eqs.~(\ref{dnc}) and (\ref{dnd}),
which have been included only to NLO. This mismatch is more evident in
the initial condition, where $\mu_{0F}$ plays a role,
since the coupling constant is larger,
being evaluated at scales that are smaller than in the
coefficient function. We should expect a milder dependence
on such a factorization scale if we matched the resummed initial condition
to the exact NNLO result \cite{alex1}. 
Moreover, the scale $\mu_{0F}$ also enters in the DGLAP evolution
operator (\ref{evol}), which we have implemented to NLL accuracy, using
NLO splitting functions. 
Ref.~\cite{alex4} has recently calculated the NNLO  corrections to the
time-like splitting functions, which include a contribution analogous to 
the one $\sim A^{(3)}$ entering in the resummed expressions (\ref{GNC}) and
(\ref{resdini}). The fact that we have not included such effects
may be a further source of mismatch, 
which may be fixed implementing the splitting functions to NNLO accuracy.
In any case, a peculiar feature of our model is that, since we are not using
any hadronization model, our theoretical error includes, at 
the same time, uncertainties of both perturbative and non-perturbative
nature. The partonic calculations in Refs.~\cite{cc,ccm,corc}, which
are NLL/NLO and use the standard coupling constant, yield indeed a
very mild dependence on the quantities which enter in the perturbative 
calculation. However, in order to predict hadron-level observables, these
perturbative computations
need to be convoluted with a non-perturbative fragmentation function, whose
parameters, after being fitted to SLD and LEP data, 
exhibit errors which are typically of the order of $10\%$, in such a way
that the hadron spectra still present fairly large uncertainties.

Turning back to Fig.~\ref{plotmu}, let
us observe that our distributions become negative and 
oscillate at very small and at very large $x_B$. 
In fact, the coefficient function (\ref{cf}) presents a term of the form
\beq
\frac{\alpha_S(\mu_R^2)\, C_F}{\pi} \, \ln x, 
\eeq
which is enhanced at small $x$ and has not been resummed in our analysis. 
In any case, that does not affect much the comparison with the data, 
since the latter are at $x>0.12$.
The coefficient function (\ref{cf}) 
and the initial condition (\ref{dbini}) also contain terms
of the type
\beq\label{log1-x}
-\,  \frac{ \alpha_S(\mu_R^2)\, C_F}{\pi} \, \ln(1-x) 
~~~~~~~ {\rm and} ~~~~~~~ 
2 \, \frac{ \alpha_S(\mu_{0R}^2)\, C_F}{\pi} \, \ln(1-x),
\eeq
which become large for $x\to 1$ and have not 
been resummed. It is at present not known how to accomplish this task.
The contributions in Eq.~(\ref{log1-x}) are responsible of
the oscillating behaviour at large $x_B$.
It is  conceivable that the inclusion of NNLO corrections to the
coefficient function and the initial condition may partly stabilize the
distribution at the endpoints. 
In any case, we are aware that our simple model, based on an extrapolation
of the perturbative behaviour to small energy scales, is expected 
to fail for very large $x$. 
It is therefore safe discarding a few data points at large $x_B$ when 
comparing with the data and, e.g., limiting ourselves to
$x_B \, \lsim \, 0.92$. 
We find that our default parameters give a good description of
the ALEPH data ($\chi^2/\mathrm{dof} \, = \, 21.4/16$), 
but reproduce rather badly
the OPAL ($\chi^2/\mathrm{dof} \, = \, 162.7/18$) and 
SLD ($\chi^2/\mathrm{dof} \, = \, 109.1/20$)
ones. The overall $\chi^2/\mathrm{dof}$, computed 
as if all measurements were coming from one
experiment  is also quite large: $\chi^2/\mathrm{dof} \, = \, 293.2/54$.
A much better description of SLD, OPAL and the overall sample is obtained
setting $\mu_{0F} \, = \, m_b/2$. We obtain: 
$\chi^2/\mathrm{dof}=$~24.8/16
(ALEPH), 30.4/18 (OPAL), 47.8/20 (SLD) and 103.0/54 (overall).
Such values of $\chi^2/\mathrm{dof}$ are
perfectly acceptable, since we are comparing data from different
experiments and, as already discussed, 
SLD and OPAL, unlike ALEPH, also reconstructed a small fraction
of $b$-flavoured baryons.

The theoretical uncertainties due to the values chosen for $\alpha_S(m_Z^2)$
and $m_b$ are explored in Fig.~\ref{plotalpha}: we consider 
$\alpha_S(m_Z^2)\, = $~0.117, 0.119 and 0.121
\footnote{The corresponding range for the
analytic time-like coupling constant is $0.115<\tilde\alpha_S(m_Z^2)<0.119$.},
and $m_b \, = \, 4.7$, $5.0$ and $5.3$ GeV. 
The impact of the choice of these quantities is comparable and well visible  
throughout all the $x_B$-range, though smaller than the one due to the
variation of $\mu_{0F}$. The effect is about 10\% at average values
of $x_B$ and grows up to 35\% at large $x$.
While changing $\alpha_S(m_Z^2)$ and $m_b$ does not improve much the comparison
with OPAL and SLD, an excellent description of the ALEPH data 
($\chi^2/\mathrm{dof}=11.9/16$)
is obtained for $m_b\, = \, 5.3$~GeV, a value compatible with $B$-meson masses,
and still $\alpha_S(m_Z^2)=0.119$.
Nevertheless, the comparison with
the other experiments gets worse for this value of the bottom-quark mass,
as we get $\chi^2/\mathrm{dof}=116.1/18$ for OPAL,  
84.2/20 for SLD and  212.2/54 overall.
As for the \msbar masses 
$\bar{m}_b$, $\bar{m}_c$ and $\bar{m}_s$, the effect of their
variation on the $x_B$-spectrum is very little: we do not show such
plots for the sake of brevity.
The most relevant values of $\chi^2/\mathrm{dof}$ when we vary the
perturbative parameters are collected in Table~\ref{tabchi}.
\begin{table}
\caption{\label{tabchi} \small Results of the comparison of our model with
LEP and SLD data, for the most significant values of $\mu_{0F}$ and
$m_b$. We have set $\alpha_S(m_Z^2)=0.119$, while 
the other parameters have very little impact.}
\vspace{0.2cm}
\begin{tabular}{| c | c | c | c | c | c |}
\hline
$\mu_{0F}$ & $m_b$ & $\chi^2/\mathrm{dof}$ (ALEPH)&
$\chi^2/\mathrm{dof}$ (OPAL)& $\chi^2/\mathrm{dof}$ (SLD)&
$\chi^2/\mathrm{dof}$ (overall)\\
\hline\hline
$m_b$&5 GeV & 21.4/16& 162.18/18& 109.1/20 & 293.2/54\\
\hline
$m_b/2$&5 GeV & 24.8/16& 30.4/18& 47.8/20& 103.0/54\\
\hline
$m_b$&5.3 GeV & 11.9/16& 116.1/18& 84.2/20& 212.2/54\\
\hline
\end{tabular}
\end{table}

\subsection{Relevance of NNLL effects}

Before turning to the analysis in moment space, we wish to estimate
the impact on our prediction of the NNLL threshold-resummation corrections.

In Fig.~\ref{plotall} we present the experimental points 
and the spectra yielded by our model using the parameters which 
give the overall best fit to the data ($\mu_{0F} \, = \, m_b/2$
and $m_b\, =\, 5$~GeV), but
within two approximations schemes: 
NNLL soft resummation along with NNLO effective coupling constant (solid line),
and NLL soft resummation with NLO coupling constant (dashed).
In the NLO/NLL prediction, we still perform the Mellin transforms of the
resummed expressions exactly, i.e. without the step-function approximation
(\ref{step}).
By dropping the NNLL terms, i.e. the contributions proportional
to $\tilde A^{(3)}$, $B^{(2)}$ and $D^{(2)}$ in the resummed expressions
(\ref{GNC}) and (\ref{resdini}), 
we see that the spectrum gets shifted to larger $x_B$, and the agreement with
the experimental data gets worse
\footnote{
Similar conclusions are reached by comparing thrust, heavy-jet mass 
and $C$-parameter distributions --- computed with the effective time-like 
coupling within NLL accuracy --- with LEP1 data 
\cite{unpub}.}. 
For $x_B \, \lsim \, 0.92$ we obtain: 
$\chi^2/\mathrm{dof} \, = \, 11.18$ (SLD), $7.01$ (ALEPH), $16.27$ (OPAL) 
and $12.05$ (overall).
The inclusion of the NNLL threshold contributions to the coefficient function 
and to the initial condition of the perturbative fragmentation function,
together with the NNLO corrections to $\tilde\alpha_S(k^2)$, is therefore
crucial to get reasonable agreement with the data and the
$\chi^2/\mathrm{dof}$ quoted in Table~\ref{tabchi}.
In fact, when using the time-like coupling constant, 
the ${\cal O}(\alpha_S^3)$ 
coefficient $A^{(3)}$ in the NNLL-resummed coefficient function and in the
initial condition gets enhanced according to Eq.~(\ref{a3}).
The change $A^{(3)}\to\tilde A^{(3)}$ shifts the peak of the spectrum towards
a lower value of $x_B$.
If we had used the space-like coupling constant $\bar\alpha_S(k^2)$, given in
Eq.~(\ref{space}), instead of $\tilde\alpha_S(k^2)$, 
we would have obtained the same $A^{(3)}$
and a rather poor description of the data.

It is also interesting to gauge the overall impact of the inclusion 
of power corrections by means of
our model. In Fig.~\ref{plotall} we also show the
NLL-resummed parton-level spectrum of Ref.~\cite{cc}, where
the standard NLO coupling constant is used, the Mellin transforms are performed
by the step-function approximation, and the inversion to $x$-space is done
using the minimal prescription \cite{min} to
avoid the Landau pole.
We see that the 
distribution of Ref.~\cite{cc} is peaked at very large $x$ and
is very far from the data for $x_B\, > \, 0.4$. 
In fact, the $b$-quark spectrum of
\cite{cc} is expected to be convoluted with a non-perturbative fragmentation
function, whose parameters need to be fitted to reproduce the data.
Our spectrum is also broader than the one of \cite{cc}: one can actually
show that this is due to the fact that we have 
performed the Mellin transforms exactly.
\begin{figure}[t]
\centerline{\resizebox{0.65\textwidth}{!}{\includegraphics{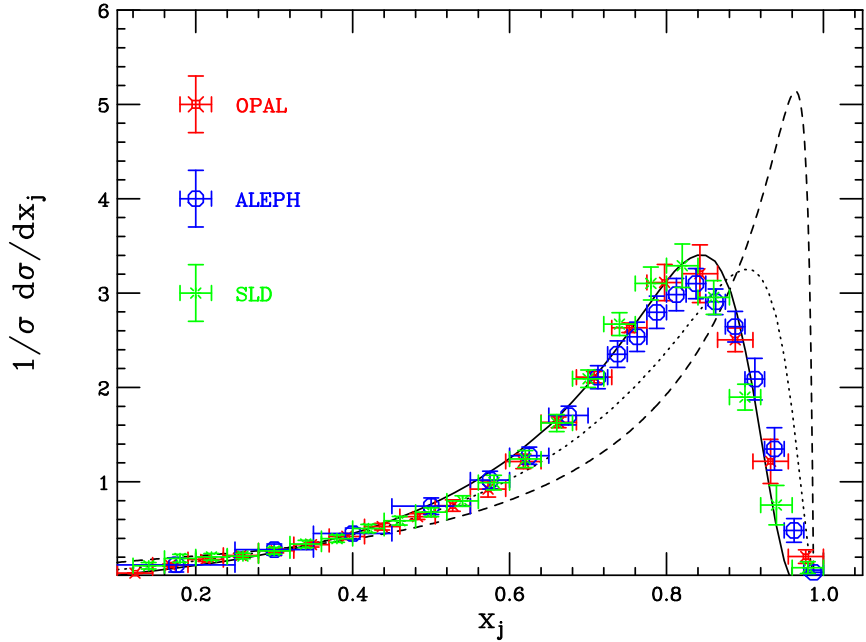}}}
\caption{\small
The solid line is our 
best prediction for $B$-hadron production ($j \, = \, B$),
obtained using NNLL threshold resummation 
and the NNLO time-like coupling, super-imposed to the data.
The dotted curve is obtained by dropping the NNLL terms, i.e. working within
NLL accuracy. 
The dashed line is the parton-level spectrum of Ref.~\cite{cc} ($j = b$), 
which uses the standard $\alpha_S(k^2)$ and NLL large-$x$ resummation. }
\label{plotall}
\end{figure}

\sect{Phenomenology --- $N$-space}
\label{sec9}

We would like to present the results yielded by our approach
in Mellin moment space, and compare them with the measurements of the
DELPHI experiment \cite{delphi}.
It was argued in \cite{canas} that working 
in $N$-space is theoretically preferable, since one does not need to 
assume any functional form for the non-perturbative fragmentation
function and can fit directly its moments. Moreover, as shown in
\cite{drol,gardi}, the moments obtained fitting a non-perturbative model
discarding data points at small and large $x_B$ are quite uncorrect, since
the tails of the distributions play a crucial role to obtain the 
right $N$-space results.
Our case is clearly different, since we are not tuning a 
non-perturbative fragmentation function, and therefore
we do not have the problems related to the fits.
However, our distributions are still negative and unreliable at very small and 
large $x_B$, and we were forced to discard few data points to obtain 
acceptable values of $\chi^2/\mathrm{dof}$.
It is therefore still
interesting to present the moments obtained using the analytic
coupling constant and explore
the theoretical uncertainties, by changing the parameters
which enter in the calculation, along the lines of our $x$-space analysis.
We calculate the
$B$ spectrum in Mellin space directly from Eq.~(\ref{sigmabhad}),
and choose the same default values
for scales, quark masses and $\alpha_S(m_Z^2)$ as in $x$-space.
\begin{table}[t]
\caption{\label{tab}\small  Moments
$\sigma^B_N$ from DELPHI~\protect\cite{delphi} and moments 
$[\sigma_N^B]_{\mathrm{th}}$ yielded by our calculation. 
We quote the uncertainties due to the parameters which enter in the
perturbative calculations, varied as discussed in the text.
The theoretical total error is estimated as
the sum in quadrature of the partial errors.
We also present the moments $\sigma_N^b$ according to Ref.~\cite{cc}.}
\vspace{0.2cm}
\begin{tabular}{| c | c c c c |}
\hline
& $\langle x\rangle$ & $\langle x^2\rangle$ & $\langle x^3\rangle$
& $\langle x^4\rangle$ \\
\hline
\hline
$e^+e^-$ data $\sigma_N^B$ & ${\bf 0.7153 \pm 0.0052}$ & 
${\bf 0.5401 \pm 0.0064}$ &
${\bf 0.4236 \pm 0.0065}$  & ${\bf 0.3406 \pm 0.0064}$  \\
\hline
\hline
$[\sigma_N^B]_{\mathrm{th}}$  & 
${\bf 0.6867 \pm 0.0403}$ & ${\bf 0.5019 \pm 0.0472}$ & 
${\bf \, 0.3815 \pm 0.0465  }$ & 
${\bf \, 0.2976 \pm 0.0462 }$ \\
\hline
$\delta\sigma_N^B(\mu_R)$ &0.0014 &0.0011 & 0.0009 & 0.0007\\
\hline
$\delta\sigma_N^B(\mu_F)$& 0.0066 & 0.0067 &0.0059 & 0.0051  \\
\hline
$\delta\sigma_N^B(\mu_{0R})$   &0.0022&0.0028  & 0.0031&0.0033   \\
\hline
$\delta\sigma_N^B(\mu_{0F})$   &0.0364 & 0.0414 & 0.0398 & 0.0364 \\
\hline
$\delta \sigma_N^B(m_b)$ &0.0111  & 0.0145 & 0.0153 & 0.0150\\
\hline
$\delta\sigma_N^B( \bar{m}_b )$
&0.0004 &0.0005&0.0006& 0.0006 \\
\hline
$\delta\sigma_N^B( \bar{m}_c)$
& 0.0003 & 0.0005 & 0.0006 & 0.0006 \\
\hline
$\delta\sigma_N^B( \bar{m}_s)$ & 0.0004 &  0.0007&0.0008 & 0.0008 \\
\hline
$\delta\sigma_N^B(\alpha_S(m_Z^2))$ &0.0113  & 0.0158 & 0.0173&0.0176  \\
\hline
$\sigma_N^b$  & 0.7734 $\pm$ 0.0232 & 0.6333 $\pm$ 0.0311 
& 0.5354 $\pm$ 0.0345 & 0.4617 $\pm$ 0.0346 \\
\hline
\end{tabular}
\end{table}
\par
In Table~\ref{tab} we present the results of our study:
we quote the experimental moments measured by DELPHI, the central
values yielded by our calculation, denoted by $(\sigma_N^B)_{\mathrm{th}}$,
and the errors due to all the quantities which we vary.
The data correspond to $N=$~2, 3, 4 and 5; the first moment is 
$\sigma_{N=1}^B=1$, 
since both data and our spectra are normalized to unity.
We estimate the overall theoretical error summing in quadrature the errors
due to the variation of each parameter.

We observe that the experimental moments exhibit very little errors and that
our central values are smaller by about $5 \div 10 \%$ than
the DELPHI ones. This result confirms what we had found in $x_B$-space:
our $x_B$-spectra go to zero more rapidly than the experimental data
for large $x_B$, and become negative for $x_B>0.96$; for low and intermediate
$x_B$, our curves lie above the data. It is therefore reasonable that 
our moments are smaller than the measured ones.
However, we find that, within the uncertainties
due to masses and scales, our calculation is in fair agreement with
the data. 
As observed for the $x_B$-spectrum, the uncertainty due to the
choice of $\mu_{0F}$ is pretty large, and the ones due to 
$\alpha_S(m_Z^2)$ and  $m_b$ are smaller but visible. The other
scales and the \msbar masses, entering in the matching of $\alpha_S$ at
different flavour numbers, have very little impact on the moments of
the $B$ cross section.
As far as the uncertainties are concerned, 
we observe that the DELPHI data are for $N \,\le \, 5$,
hence the terms $\sim\ln^k N$ in the resummed exponents
are not so dominant with respect to other
contributions, such as the constants.
Therefore, even in $N$-space, we expect
that the further inclusion of NNLO contributions will have a
significant impact on the prediction and decrease the 
theoretical error.

We also present in Table~\ref{tab} the moments yielded by the NLL-resummed
calculation of \cite{cc}, which uses the NLO standard $\alpha_S(k^2)$, whose
$x$-space results have already been displayed in Fig.~\ref{plotall}.
We estimate the error on the moments of Ref.~\cite{cc} as we did for the
one on our calculation, varying the perturbative parameters in the same range.
We find that the $N$-space results of \cite{cc},
denoted by $\sigma_N^b$ 
to stress that they are a parton-level result,
are very far from the data and much larger than 
the moments obtained using the analytic 
coupling constant. This result
is in agreement with the plots in Fig.~\ref{plotall}, and confirms
the remarkable impact of non-perturbative effects in $N$-space as well.
Moreover, we observe that the errors on the moments of \cite{cc} are
smaller than the ones yielded by our calculation. In fact, 
Ref.~\cite{cc} employs NLL threshold resummation and DGLAP
evolution, and matches the resummation to the NLO results. As already 
discussed in the $x$-space analysis, using NNLL large-$x$ resummation 
but still NLL DGLAP evolution and NLO remainder functions, as we did,
may lead to a
mismatch which produces larger uncertainties. We expect that
the inclusion of NNLO corrections to the initial condition and to the
splitting functions will lead to a weaker dependence on the
input parameters of our analysis. In any case, the perturbative moments
of \cite{cc} need to be multiplied by the moments of a 
non-perturbative fragmentation function extracted from the data, which
will lead to larger uncertainties on the hadron-level moments.

\sect{Conclusions}
\label{sec10}

We studied $B$-hadron energy distributions in $e^+e^-$ annihilation at the
$Z^0$ peak by means of a model having as the source of
non-perturbative corrections an effective QCD coupling constant.

The physical idea behind our model is that the main non-perturbative
effects, related to soft interactions in hadron bound states, are not very 
strong coupling phenomena, but can be described
by an effective coupling constant 
of intermediate strength, typically $\tilde{\alpha}_S \, \approx \, 0.3-0.5$.
That implies that, within our model, these non-perturbative corrections ---
the Fermi motion in $B$ decays being the best-known example ---
can be obtained from perturbation theory by means of an extrapolation.

We modelled power corrections using a NNLO effective 
$\tilde{\alpha}_S(k^2)$ constructed removing first the Landau pole,
according to an analyticity requirement, and then resumming to
all orders the absorptive effects related to time-like gluon 
branching. The resulting
$\tilde{\alpha}_S(k^2)$ significantly differs from the standard 
$\alpha_S(k^2)$ at small scales.
We described $b$-quark production 
within the framework of perturbative fragmentation functions,
resumming threshold logarithms in the coefficient function 
and in the initial condition of the perturbative fragmentation function
to NNLL accuracy, and matching the resummed results to the NLO ones.
The perturbative fragmentation function was evolved using
NLL DGLAP equations.
When implementing threshold resummation in $N$-space, as part of
our model, we decided to perform the Mellin transforms exactly, which leads
to the further inclusion in the Sudakov exponent of
terms $\sim\ln^k N$ of higher order (N$^3$LL, N$^4$LL, $\dots$), 
along with some constants and 
power-suppressed terms ${\cal O}\left( 1/N \right)$.

We presented results on the $B$-hadron energy distribution, 
relying on this model to include power corrections, 
without introducing any non-perturbative fragmentation function
with tunable parameters.
When studying the $B$ spectra, we investigated the dependence 
of our prediction on the quantities 
which enter in the perturbative calculation, such as quark masses, 
renormalization and factorization scales, and $\alpha_S(m_Z^2)$.
We observed that the $x_B$-distributions become negative and oscillating
at very small and large $x_B$, which is due 
terms $\sim\ln x$ and $\sim\ln(1-x)$, that are present in the NLO
remainder functions and have not been resummed.
We have therefore discarded a few data points at large $x_B$, where
our calculation is anyway unreliable, and limited ourselves to 
$x_B\lsim 0.92$.
We compared our results with OPAL, ALEPH and
SLD data on $b$-flavoured hadron production and found that, within our
theoretical uncertainties, our calculation is able to reproduce quite well
the ALEPH and OPAL data, while it is marginally consistent with SLD.
As for the dependence on the perturbative parameters, 
the effect of the choice of the factorization scale 
$\mu_{0F}$, appearing in the initial condition of the perturbative 
fragmentation function, is quite large and the data are better described 
for low values of $\mu_{0F}$. The dependence on $\alpha_S(m_Z^2)$ and on
$m_b$ is also quite visible: for example, a value of $m_b$ consistent with
$B$-meson masses, which is reasonable since our model assumes
$E_B \, \simeq \, E_b$, yields an excellent description of the ALEPH data.
The other parameters have instead very little impact on the
$x_B$-spectrum.

We also compared our results in Mellin space with the moments 
measured by the DELPHI collaboration. We found that the central values 
of our $N$-space results are smaller than the experimental ones,
but nonetheless, within the theoretical uncertainties, 
our moments are consistent with the data. 
In particular, a fairly large uncertainty is still due to
the choice of $\mu_{0F}$, as was observed in the $x$-space analysis.

In summary, we find it remarkable that, within our theoretical uncertainties,
we have been able to describe
the LEP data in both $x$- and $N$-spaces,
by modelling non-perturbative corrections via the
analytic time-like coupling constant and without tuning any parameter to
such data. It is also pretty interesting that a model which succeeded
in reproducing the photon- and hadron-mass energy distribution in
$B$-meson decays \cite{noi4} has led to a reasonable fit of
$B$-production data, although they are quite different processes,
characterized by different energy scales.

Our model looks therefore quite promising and we plan to further apply
it to other processes and observables. A straightforward extension of the
study here presented is the investigation
of $B$ production in top and Higgs decays, using the perturbative
calculations in \cite{ccm,cm,corc}, and modelling power corrections as
in the present paper. We can also use our model along with Monte Carlo
event generators: in Ref.~\cite{drol}, the hadronization models of
HERWIG and PYTHIA were tuned to the same data as the ones here considered, 
and then used to predict $B$ production in top and Higgs decays.
We may thus think of using the parton shower algorithms of HERWIG
and PYTHIA to describe perturbative $b$-production,
with our analytic coupling constant in place of the cluster and
string models. At the end of the cascade, we can assume $E_b\to E_B$ and
compare the results with the data.  
The spectra yielded by Monte Carlo
event generators are positive definite and do not exhibit the problem of 
becoming negative. However, parton shower algorithms are equivalent
to a LL/LO resummation, with the further inclusion of some
NLLs \cite{cmw}. Since we have learned from this analysis that the inclusion
of corrections of higher orders is crucial to reproduce the data, we may have
to add to the Monte Carlo Sudakov form factor
contributions analogous to the NLLs and NNLLs that are missing,
in particular the one $\sim \tilde A^{(3)}$.

Besides, it will be really worthwhile to use the recent calculations
of the NNLO initial condition of the perturbative fragmentation function
\cite{alex1,alex2} and of the NNLO splitting functions in the non-singlet 
sector 
\cite{alex3} to fully promote our formalism to NNLO/NNLL accuracy.
This way we could explore whether the uncertainties on our predictions 
--- especially the one due to the scale $\mu_{0F}$ ---
get finally reduced.
If the theoretical uncertainties get substantially reduced, 
one may even think
of extracting $\alpha_S(m_Z^2)$ from $b$-fragmentation data, as already done
from $B$-decay spectra \cite{noi4}.

Our study could also be improved by including NNNLL contributions in
the resummation exponents. Such corrections involve the coefficients
$A^{(4)}$, $B^{(3)}$ and $D^{(3)}$: at present only $B^{(3)}$ 
is exactly known.
In particular, we expect a relevant effect due to the possible inclusion of 
$D^{(3)}$, the ${\cal O}(\alpha_S^3)$
coefficient of function $D(\alpha_S)$, which resums large-angle soft
radiation in the initial condition.
In fact, we have observed that the scale of $\alpha_S$ in the
resummed initial condition is smaller than in the coefficient function,
and therefore power corrections are more relevant.
Also, our model enhances the coefficients of $\alpha_S$ from the
third order on. We can therefore employ our model to 
implement non-perturbative corrections to Drell--Yan or
Deep Inelastic Scattering processes,
whose coefficient function was recently calculated to NNNLO accuracy
\cite{dis}. 
In $e^+e^-$ annihilation, threshold contributions to the 
NNNLO coefficient function were computed in \cite{ravi}; the
full ${\cal O}(\alpha_S^3)$ corrections have not been calculated yet.

From the theoretical viewpoint, we plan to investigate in more detail
the issue of the Mellin transform of the resummed cross section and
the power corrections which are inherited by the $B$ spectrum when
one uses the effective $\tilde\alpha_S(k^2)$ and
does the longitudinal-momentum integration in an exact or approximated way.
Within our model, we chose to perform it exactly, driven by the 
results in \cite{noi4,Aglietti:2004fz}, but nonetheless we believe 
that a thorough study of this point, along the lines of
\cite{min,braun}, should be necessary. 

Other issues which we plan to investigate in detail are
the treatment of the higher orders of the effective $\tilde\alpha_S$ 
and the comparison between time- and space-like coupling constants.
We have obtained reasonable agreement with the data by using the 
time-like effective coupling constant and
taking the powers $\tilde\alpha_S^n(k^2)$ after performing the
integral of the dispersion relation.
Nonetheless, this conclusion is somehow related to the level of approximation
of the perturbative calculation which we have employed, and may not hold
if we used a different accuracy.
Therefore, a more solid understanding of these aspects of our 
model is mandatory.

Furthermore, 
our model has some built-in universal features implying that, to put it into
a stringent check, one should consider several observables from 
different processes.
To this goal, however, one might need both NNLL resummation
from the theoretical side and accurate data on the experimental one.
Shape-variable data at the $Z^0$ peak, for example, are rather accurate, 
but for these quantities NNLL resummation is still in progress, 
hence preventing an analysis within our model.
In \cite{noi4} this model was applied to semi-inclusive $B$ decays, where
the situation is somehow complementary with respect to 
shape variables in $e^+e^-$ annihilation: 
NNLL threshold resummation is well established, 
but data are not very accurate yet, because of large backgrounds.

Finally, we could also push our model to the `low-energy' direction, 
to try to 
describe, for example, charm production at the $Z^0$ peak or below it, 
where accurate experimental data are available.
Due to the large value of $\alpha_S(m_c^2)\simeq 0.35$
and to a soft scale $S\simeq m_c(1-x)$, 
smaller by a factor of three than in $b$-production,
we expect a full NNLL/NNLO analysis to be necessary.
The comparison with $D$-hadron data will be crucial to investigate possible
deviations from our model.
In fact, an extension of the formulation here presented may
consist in adding to
$\tilde\alpha_S(k^2)$ a correcting term:
$\tilde\alpha'_S(k^2)= \tilde\alpha_S(k^2)+
\delta\tilde\alpha_S(k^2)$.
This way, $\tilde\alpha_S(k^2)$ will still be the effective
coupling constant here discussed, while 
$\delta\tilde\alpha_S(k^2)$ will depend on the hadronic state which we wish to
describe, and vary, e.g,  
from mesons to baryons or from $B$'s to $D$'s. 
Investigating the charm sector may therefore help in shedding 
light on our model and understanding whether such a correcting term is
necessary or not. This study is in progress as well. 

\vspace{0.5cm}
\centerline{{\large \bf Acknowledgments}}

\noindent
We are indebted to S.~Catani for discussions on soft-gluon resummation.
We also acknowledge M.~Cacciari for discussions on the perturbative
fragmentation approach and for providing us with the computing
code to obtain the
results of Ref.~\cite{cc}.

\appendix
\sect{Numerical evaluation of Mellin transforms}
In this appendix we present a method for computing numerically the  
Mellin transform, through the Fast Fourier Transform (FFT), of a function 
of the form
\beq
f(y) \, = \, \left[ \frac{\varphi(y)}{y} \right]_+ ,
\eeq
where $y \, = \, 1-z$ and $\varphi(y)$ is a regular
function of $y$ or a function having at most a logarithmic singularity
for $y \, \to \, 0$.

To calculate the integer moments of the cross section $\sigma_N$
--- typically the first few moments with $N = 2,3,4,5,\cdots$ ---
the integral defining the Mellin transform does
not present any convergence problem and can be done directly.
In order to obtain the cross section in $x$-space, it is however 
necessary to perform an inverse Mellin transform by integrating 
$\sigma_N$ along a vertical line in $N$-space.
The numerical computation of $\sigma_N$ in the
complex $N$-plane is non trivial for 
${\rm Im}\, N \, \gg \, 1$ because the kernel $z^{N-1}$ develops 
fast oscillations with $z$, which affect the convergence of
the integral.

In detail, we have to compute the integral
\beq
\label{g_N}
g_N \, = \, \int_0^1 dy (1-y)^{N-1} \, f(y)
\, = \, \int_0^1 \frac{dy}{y} \, \left[ (1-y)^{N-1} - 1\right] \, \varphi(y).
\eeq
for $N$ lying on a vertical line in the complex plane,
\beq
N \, = \, c \, + \, i \nu,
\eeq
with $c \, > \, 0$ and $\nu$ real.
It is convenient to treat analytically the infrared cancellation
between real and virtual contributions related to the first and
the second term in square brackets on the r.h.s. of Eq.~(\ref{g_N}).
We then take a derivative with respect to $\nu$:
\beq
\gamma_c\left( \nu \right) 
\, \equiv \, \frac{d g_{c + i \nu} }{d \nu}
\, = \, 
i \int_0^1 dy \, \frac{\ln(1-y)}{y} \, (1-y)^{c-1 + i \nu} \, \varphi(y).
\label{gammac}
\eeq
Note 
that the infrared singularity  $1/y$ is now regulated by the $\ln(1-y)$ factor
coming from the differentiation.

In order to express $\gamma_c(\nu)$ as the Fourier transform (FT)
of some function, we change variables to
\beq
\label{def_t_y}
t \, \equiv \, - \, \ln(1-y),
\eeq
and express Eq.~(\ref{gammac}) as an integral over $t$:
\beq
\gamma_c\left( \nu \right) 
\, = \, 
\int_0^{\infty} dt \, e^{- \, i \, \nu \, t} \, \psi_c(t),
\eeq
with
\beq
\psi_c(t) \, \equiv \, - i \, \frac{t}{1 \, - \, e^{- t} } \, e^{- c \, t} 
\, \varphi\left[ 1 \, - \, e^{- t} \right].
\eeq
$\gamma_c$ is therefore the Fourier transform of the function $\psi_c(t)$
defined above and vanishing for $t \, < \, 0$.
For a fast numerical evaluation, it is convenient to transform the
FT above into a Fast-Fourier-Transform (FFT).
We cut the improper integral at a large but finite $t_{\max}$:
\beq
\gamma_c\left( \nu \right) 
\, \simeq \, 
\int_0^{t_{\max}} dt \, e^{ - \, i \, \nu \, t } \, \psi_c(t).
\eeq
In practice, because of the exponential dependence $y\,\approx \, \exp[-t]$,
we found that 
$t_{\max} \, = \, 20 \div 30$ already gives a good accuracy.
The above integral can be easily approximated by means of a constant
sampling:
\beq
\Delta t \, \equiv \, \frac{ t_{\max} }{n},
\eeq
where $n$ is the number of points in which the function $\psi_c(t)$
is evaluated. We have found that an accuracy $O(10^{-3}\div 10^{-4})$ 
is reached 
already with $n \,= \, 10^4 \div 10^5$ points.
We then have:
\beq
\gamma_c\left( \nu \right) 
\, \cong \,
\, e^{ - \, i/2 \, \nu \, \Delta t }\,\Delta t
\, \sum_{k=0}^{n-1}  e^{ - i \, \nu \, k \, \Delta t }
\, \psi_c\left[ \left( k + \frac{1}{2}\right) \Delta t\right].
\eeq
The Mellin transform is obtained by a numerical integration of $\gamma_c(\nu)$:
\beq
\label{mellin_fin}
g_{c + i \nu} \, = \, g_c \, + \, \int_0^{\nu} d\nu' \, \gamma_c\left( \nu' \right). 
\eeq
$g_c$ is the Mellin transform in the fixed point $c$ on the
positive $N$ axis; it is a constant which is computed directly
just once.
Eq.~(\ref{mellin_fin}) is our final result for the numerical computation
of the Mellin transform.

Let us now discuss the numerical computation of the inverse Mellin transform
with a similar method:
\beq
f(y) \, = \, \int_{c-i\infty}^{c+i\infty} \frac{dN}{2\pi i} \, (1-y)^{-N}
 \, g_N
\, = \, 
\frac{1}{(1-y)^c} \,
\int_{-\infty}^{+\infty} \frac{d\nu}{2\pi} \, (1-y)^{-i \nu} \, g_{c+i \nu}.
\eeq
Since $f(y)$ is real, one immediately obtains the following property of
its Mellin transform:
\beq
\left( g_N \right)^* \, = \, g_{N^*}.
\eeq
The inverse Mellin transform can therefore also be written as:
\beq
\label{mellin_inv}
f(y) \, = \, 
\frac{1}{\pi \, (1-y)^c} \,
{\mathrm {Re}} \, \int_0^{\infty} d \nu \, 
e^{ - \, i \, \nu \, \ln (1-y) } \, g_{c \, + \, i \, \nu}.
\eeq
Apart from constant factors,
Eq.~(\ref{mellin_inv}) express the inverse Mellin transform as the
inverse Fourier transform $\nu \, \to \, - \, \ln (1-y)$ of $g_{c + i \nu}$.
The transformation to the Inverse Fast Fourier Transform can be made
as in the direct case.

We have implemented the above algorithm within the Mathematica System
with a gain of CPU time by over an order of magnitude with respect to 
the direct numerical evaluation.
A typical run on a standard PC takes ${\cal O}(1)$ minute.
We have also checked our numerical results with direct
integration in Fortran.

\end{document}